\documentclass[aps,nofootinbib,preprintnumbers]{revtex4-1}

\usepackage{amsmath,amssymb,amsfonts,graphicx,float}
\usepackage{color}
\usepackage{subfigure}
\usepackage[hidelinks]{hyperref}

\newcommand{\beq}{\begin{eqnarray}}
\newcommand{\eeq}{\end{eqnarray}}

\newcommand{\tabii}{\hspace{.2\textwidth}}

\begin{document}

\title{Power-law Optical Conductivity from Unparticles:  Application to the Cuprates}
\author{Kridsanaphong Limtragool}\author{Philip Phillips}\thanks{Guggenheim Fellow}
\affiliation{Department of Physics and Institute for Condensed Matter Theory,
University of Illinois
1110 W. Green Street, Urbana, IL 61801, U.S.A.}

\date{\today}

\begin{abstract}

We calculate the optical conductivity using several models for unparticle or scale-invariant matter. Within a Gaussian action for unparticles that is gauged with Wilson lines, we find that the conductivity computed from the Kubo formalism with vertex corrections yields no non-trivial deviation from the free-theory result.  This result obtains because at the Gaussian level, unparticles are just a superposition of particle fields and hence any transport property must be consistent with free theory.  Beyond the Gaussian approach, we adopt the continuous mass formulation of unparticles and calculate the Drude conductivity directly.  We show that unparticles in this context can be tailored to yield an algebraic conductivity that scales as $\omega^{-2/3}$ with the associated phase angle between the imaginary and real parts of $\arctan\frac{\sigma_2}{\sigma_1}=60^\circ$ as is seen in the cuprates.   Given the recent results\cite{Donos2014,Rangamani2015,Langley} that gravitational crystals lack a power-law optical conductivity, this constitutes the first consistent account of the $\omega^{-2/3}$ conductivity and the phase angle seen in optimally doped cuprates.  Our results indicate that at each frequency in the scaling regime,  excitations on all energy scales contribute.  Hence, incoherence is at the heart of the power-law in the optical conductivity in strongly correlated systems such as the cuprates. 

\end{abstract}

\maketitle

\section{Introduction}
Among the signatures of non-Fermi liquid behavior, the fractional power law observed in the optical conductivity of optimally doped cuprates, $\sigma(\omega) \propto \omega^{-\alpha}$\cite{Marel2003,Basov2011} with  $\alpha=2/3$, stands as a key failure of the standard theory of metals. This power law is ubiquitous, having been observed first in  YBa$_2$Cu$_3$O$_{6+\delta}$ \cite{Schlesinger1990,ElAzrak1994} with $\alpha = 0.5 - 0.7$ and more extensively in  Bi$_2$Sr$_2$CaCu$_2$0$_{8+\delta}$  \cite{Marel2003,Hwang2007} in the mid-infrared frequency range ($\omega$ = 500 $cm^{-1}$ to 5000 $cm^{-1}$).  Consistent with this power law is the observed phase angle of $\pi\alpha/2$\cite{Marel2003}. Realizing that some sort of scale invariant entities must be relevant, Anderson\cite{Anderson1997} computed the mid-infrared power law in the cuprates using Luttinger liquid Green functions.   He argued that in the holon nondrag regime, the vertex term can be dropped out of the conductivity calculation, $\sigma(\omega) \sim \frac{1}{\omega}\int dx\int dtG^e(x,t)G^h(x,t)e^{i\omega t}$, where $G^e$ and $G^h$ are the electron and hole Green functions of a 2D Luttinger liquid, respectively.   Although no specific calculation was provided, he posited that the Green function of 2D Luttinger liquids (an unknown state of matter) is identical to its 1D counterpart. Using the scaling properties of the product of Green functions, Anderson found  the optical conductivity to have a power law of a form, $\sigma(\omega) \propto (i\omega)^{-1+2a}$ where $a$ is the Luttinger parameter. 

Given that power laws are the finger print of criticality, Anderson's starting point of scale-invariant excitations\cite{Anderson1997} is certainly reasonable.  However since there is no basis for using 1D Luttinger liquids in 2D, it is advantageous to resort to a relatively new tool which is capable of describing quantum critical matter in higher dimensions.   Previously, one of us adopted\cite{Phillips2013} the unparticle picture of Georgi's\cite{Georgi2007a} as a model for treating the breakdown of the particle picture in the normal state of the cuprates.  The key idea here is that scale invariance places severe restrictions on the form of the propagator that are completely dimension independent and hence obviates the need to tether the description to 1D Luttinger-liquid theory.  We investigate two scale invariant models here.  In the first model, we use a Gaussian form for the unparticle action as a basis for calculating the optical conductivity. We apply the vertex couplings of scalar unparticle and gauge fields introduced in Refs. \cite{Terning1991, Cacciapaglia2008} and linear response theory at finite temperature. We show that the conductivity of unparticles is a constant factor multiplied by the conductivity of free massless scalar fields. In the second model, we turn to the continuous mass formalism that one of us originally used to show that the action for unparticles is identical to an action on anti de Sitter space\cite{Phillips2013}. This same model has been recently adopted\cite{Karch2015} to obtain the hyperscaling violation exponent and anomalous dimension for the current.  Within this approach, we obtain a power law in optical conductivity $\sim \omega^{-\alpha}$ and the associated phase angle of $\pi\alpha/2$. The implication of this result for anomalous dimensions for the current\cite{Karch2015} are discussed.

\section{Unparticle}

Although one of us explicitly argued for the relevance of unparticles to the pseudo gap phase of the cuprates elsewhere\cite{Phillips2013}, we summarize the key claim here for completeness. Numerous arguments have been presented for quantum critical points occurring in somewhere in the pseudo gap phase of the cuprates\cite{Marel2003,mfl}.  However, precisely what are the propagating degrees of freedom has proven elusive to enumerate because the underlying system is strongly correlated. Ref. \cite{Phillips2013}'s arguments for the relevance of unparticles to the pseudo gap phase focused on the existence of Fermi arcs. Ref. \cite{Phillips2013} argued that because Fermi arcs require the presence of zeros of the single-particle Green function either at the terminus of the arc or at the backside of the arc, the excitations that give rise to them arise from a divergent self energy and hence are not connected adiabatically to a Fermi liquid.  In general the new excitations are some composites that are  determined by collective phenomena.  It is precisely composite excitations that conformal field theory (CFT) attempts to describe as the relevant degrees of freedom at critical points.   Since any non-trivial infrared dynamics in strongly correlated electron systems must be controlled by a critical fixed point, scale invariance can be used to construct the form of the underlying propagator.  It is this motivation that led Georgi\cite{Georgi2007a,Georgi2007b} to forgo the full machinery of conformal field theory and just use scale invariance to construct the form of the relevant IR propagator. One of us put this proposal to use\cite{Phillips2013} in condensed matter systems because it is valid in any dimension and hence permits immediate progress beyond the restrictive 1D Luttinger liquid paradigm advanced by Anderson\cite{Anderson1997}.  

Regardless of the explicit mathematical machinery (multi flavors, summation over masses, for example) used to construct unpaticles, unparticle stuff\cite{Georgi2007a,Georgi2007b} is a scale invariant sector that emerges in the infrared (IR) of an effective field theory.  Unparticles can be thought\cite{UVIR} to arise from a general field theory in which integrating out the ultra-violet (UV) sector generates new degrees of freedom at low energies, that is, UV-IR mixing.  The signature of such mixing is an interpolating field that now resides in the low-energy sector\cite{UVIR} as in the charge $2e$ boson in the Hubbard model\cite{Phillips2013}, the basic model used to describe the cuprates.
This bosonic field mediates new electronic states at low energy that have no overlap with bare electrons and hence gives rise to zeros  of the single-particle Green function, thereby signaling the breakdown of the Landau Fermi-liquid picture.  We can understand the nature of the excitations that emerge by characterizing the new fixed point.  In principle it is difficult to establish the existence of fixed points in strongly correlated systems.  However, since all fixed points possess scale invariance, we can write the candidate propagators down immediately.  The propagator of a scalar unparticle in $d+1$ Lorentzian space-time dimension can be written immediately from scale invariance as
\beq
G(k) &=& \frac{A_{d_U}}{2\sin(d_U\pi)}\frac{i}{(-k^2-i\epsilon)^{\frac{d+1}{2}-d_U}}
\eeq
with
\beq
A_{d_U} &=& \frac{16\pi^{5/2}}{(2\pi)^{2d_U}}\frac{\Gamma(d_U+\frac{1}{2})}{\Gamma(d_U-1)\Gamma(2d_U)}.
\eeq
The unparticle scaling dimension $d_U$ must be greater than $\frac{d-1}{2}$ as required by the unitarity bound. When $d_U$ is not an integer, there are no poles in the propagator $G(k_0,\vec{k})$ when considering the complex variable $k_0$ . Instead, there are singular points at $k_0 = \pm \vec{k}^2$ with a branch cut coming out from each point. In general, a single quasi-particle excitation is associated with a first order pole in the propagator. Hence,  unparticle stuff is a theory with no single quasi-particle excitations. We introduce below two formalisms of unparticles that we use to calculate the optical conductivity. Indeed a well known example of unparticles in condensed matter systems is that of a Luttinger liquid.  However, this is valid only in $d=1$.  A key result one of us also showed in the original paper on unparticles in condensed matter system\cite{Phillips2013} is that the continuous mass formulation of unparticles is equivalent to an action in an anti de Sitter  (AdS)spacetime.   That is, the summation over mass (or flavors) is equivalent to a summation over energy and hence the connection with the AdS radius.  Hence, unparticles receive contributions from all energy scales. That is they represent the incoherence, the identifying feature of doped Mott insulators.  Consequently, their relevance to cuprate physics is quite natural.

\subsection{Unparticle Effective Action}

Unless we have a specific conformal field theory model (e.g. see Ref \cite{Georgi2008,Georgi2010}), we do not know how the unparticle field couples to the $U(1)$ gauge field and how to write down the corresponding four-point correlation function. Both gauge coupling and four-point correlation functions are necessary to compute the optical conductivity.   To this end, we write down a specific model with a Gaussian action,
\beq \label{eq:action}
S = \int \frac{d^{d+1}p }{(2\pi)^{d+1}}\ \phi^\dagger(p)iG^{-1}(p)\phi(p)
\eeq
where $\phi(p)$ is a complex unparticle field in momentum space. This effective action can be used to correctly reproduce the two-point correlation function. One must keep in mind that this action is a particular model of unparticles in which the forms of higher-point correlation functions are specified by Wick's theorem. An implication of this form of action is that an unparticle field is treated as if it is a free particle propagating with an algebraic form of momentum. We show how one can gauge this action and use it to obtain the conductivity in section \ref{sec:optical_action}.  Nonetheless, the inherent particle content of this description leads to an underlying free form for the conductivity.  This calculation is also instructive because it illustrates why the Luttinger liquid description of Anderson's\cite{Anderson1997} cannot ultimately yield a power-law for the optical conductivity.

\subsection{Continuous Mass Formalism} 

\label{sec:mass_sum}
Unparticle stuff can be formulated as an integration over the mass of free fields \cite{cm1,Deshpande2008}.   The general principle underlying this statement is that a scale invariant theory can be generated from a summation over the large number of flavors or from an integration over mass. Let $\phi_i$ be a free scalar field of flavor $i$ with mass $m_i$, ranging from 0 to infinity, and let the density of flavors be $f(m^2)$. The number of fields between $m^2$ to $m^2+dm^2$ is $f(m^2)dm^2$. An unparticle field can be defined as
\beq
\phi_U(x) \equiv \sum\limits_{i} \phi_i(x) = \int\limits_{0}^{\infty}dm^2 \ f(m^2)\phi(x,m^2).
\eeq
We will see that this definition is not the most general if unparticles serve to describe incoherent electronic states.
As shown in Ref. \cite{Deshpande2008}, the unparticle two-point correlation function is given by
\beq
G(p) = \int d^{d+1}x e^{ip\cdot x}\langle \phi_U(x)\phi_U(0) \rangle = \int\limits_{0}^{\infty} dm^2 \frac{i}{p^2 - m^2 + i\epsilon} f^2(m^2). \label{eq:pro_mass_sum}
\eeq
For a choice of $f(m^2) \sim (m^2)^{(d_U - \frac{d+1}{2})/2}$, the two-point correlation function is
\beq
G(p) \sim \frac{i}{(-p^2-i\epsilon)^{\frac{d+1}{2}-d_U}}
\eeq
The interpretation of this formalism is that an unparticle field is an effective excitation (or effective field) of a continuum of free scalar fields with varying masses. The continuous mass formulation description was further generalized in Ref. \cite{Phillips2013} by introducing mass as another coordinate in a scalar field Lagrangian. The action can then be recast into Poincar\'{e} coordinates of anti de Sitter space. Using the AdS/CFT correspondence, Ref. \cite{Phillips2013} calculated the two-point correlation function on the boundary theory and identified the unparticle propagator. Ref. \cite{Karch2015} showed that the hyperscaling violation exponent and anomalous dimension for the current can be generated by summing over the free energy in a multi-band system. Such an idea is completely equivalent to the continuous mass formulation and hence we will use the two approaches interchangeably here. Instead of calling it a multi-band system as in Ref. \cite{Karch2015}, we say that our system has a large number of flavors of free fields. We apply the continuous mass formalism to show how one can get a power law in the conductivity in a system with a large number of flavors (or multi-band system) in section \ref{sec:optical_sum}. 

\section{Optical Conductivity from Unparticle Effective Action} 
\label{sec:optical_action}
In order to calculate a response function to an electromagnetic field, we need to know how the unparticle is coupled to a U(1) gauge field. To this end, we follow the approach for gauging a non-local Lagrangian with a Wilson line from Refs. \cite{Mandelstam1962, Terning1991}. This approach was applied to scalar unparticles in Ref. \cite{Cacciapaglia2008} and to unfermion in Ref. \cite{Galloway2009}. To gauge the action (Eq. \ref{eq:action}), one rewrites it in position space,
\beq
S = \int d^{d+1}xd^{d+1}y \ \phi^\dagger(x)F(x-y)\phi(y)
\eeq
where $F(x-y)$ is a function resulting from converting the action to position space. The action can be gauged by inserting the Wilson line, $W(x,y) = exp(-ig\int\limits_{x}^{y}dw^\mu A_\mu)$, between $\phi^\dagger(x)$ and $\phi(y)$ with $g$ being charge of unparticle. The new action,
\beq
S = \int d^{d+1}xd^{d+1}y \ \phi^\dagger(x)F(x-y)W(x,y)\phi(y),
\eeq
is now $U(1)$ invariant. The vertex couplings of the unparticle and gauge fields can be obtained by taking derivatives of the action with respect to the fields. For the calculation of the optical conductivity, we need the vertex of two unparticle fields with one gauge field,
\beq
g\Gamma^\mu(p,q) &=& \frac{\delta^3S}{\delta A^\mu(q) \delta\phi^\dagger(p+q) \delta\phi(p)} \nonumber \\
&=& g(2p^\mu+q^\mu)\mathcal{F}(p,q)
\eeq
and the vertex of two unparticle fields with two gauge fields,
\begin{align}
g^2\Gamma^{\mu\nu}(p,q_1,q_2) = & \frac{\delta^4S}{\delta A^\mu(q_1) \delta A^\nu(q_2) \delta\phi^\dagger(p+q_1+q_2) \delta\phi(p)} \nonumber \\
= & g^2\bigg(2\eta^{\mu\nu}\mathcal{F}(p,q_1+q_2)  +\frac{(2p+q_2)^\nu(2p+2q_2+q_1)^\mu}{q_1^2+2(p+q_2)\cdot q_1}(\mathcal{F}(p,q_1+q_2)-\mathcal{F}(p,q_2)) \nonumber \\
& \tabii +\frac{(2p+q_1)^\mu(2p+2q_1+q_2)^\nu}{q_2^2+2(p+q_1)\cdot q_2}(\mathcal{F}(p,q_1+q_2)-\mathcal{F}(p,q_1))\bigg).
\end{align}
The function $\mathcal{F}(p,q)$ is given by
\beq
\mathcal{F}(p,q) = \frac{iG^{-1}(p+q) - iG^{-1}(p)}{(p+q)^2-p^2}.
\eeq
These vertices satisfy the Ward-Takahashi identities,
\beq
-iq_{\mu}\Gamma^{\mu}(p,q) = G^{-1}(p+q)-G^{-1}(p)
\eeq
and
\beq
q_{1\mu}\Gamma^{\mu\nu}(p,q_1,q_2) = \Gamma^\nu(p+q_1,q_2) - \Gamma^\nu(p,q_2).
\eeq
The thermal version of the unparticle propagator and the vertex couplings, both of which are needed for the optical conductivity calculation, are shown in the Appendix \ref{app:unparticle_matsubara}. The outline of the calculation is as follows. First, we compute a Fourier component of an imaginary time response function to an electromagnetic field from
\beq  \label{eq:linear_response}
K^{\mu\nu}_{-n,-n'}(-q,-q') = -\frac{(2\pi)^{2d}}{T^2}\mathcal{Z}^{-1}\frac{\delta^2}{\delta A_{\mu,n}(q)\delta A_{\nu,n'}(q')}\bigg\vert_{A=0}\mathcal{Z}[A],
\eeq
where $\mathcal{Z[A]} = \int D\phi D\phi^\dagger e^{-S[A]}$ is a partition function of the action given in Eq. \ref{eq:unparticle_action}. The subscripts, $n$ and $n'$, in the Fourier component of the response function above denote the dependence on the bosonic Matsubara frequency ($\omega_m = \frac{2\pi m}{\beta}$ with $m$ being an integer). In a system with translational invariance, the response function satisfies the relation
\beq \label{eq:trans_inv}
K^{\mu\nu}(\tau_1,x_1;\tau_2,x_2) = K^{\mu\nu}(\tau_1-\tau_2,x_1-x_2).
\eeq
Fourier transforming Eq. \ref{eq:trans_inv}, one finds
\beq \label{eq:response_translational}
K^{\mu\nu}_{n,n'}(q,q') = \frac{(2\pi)^d}{T}\delta^d(q+q')\delta_{n+n',0}K^{\mu\nu}_{n}(q).
\eeq
We use Eq. \ref{eq:response_translational} to reduce from the Fourier component of $K^{\mu\nu}(\tau_1,x_1;\tau_2,x_2)$ to that of $K^{\mu\nu}(\tau_1-\tau_2,x_1-x_2)$. After computing the response function, $K^{\mu\nu}_{n}(q)$, one obtains the conductivity in Matsubara space from
\beq \label{eq:sigma_K}
\sigma^{\mu\nu}(i\omega_n) = \lim\limits_{q\rightarrow0}\frac{1}{\omega_n}K^{\mu\nu}_n(q).
\eeq
Here we take the the thermodynamic limit $q\rightarrow0$. Finally, we perform an analytic continuation from the Matsubara to real frequency. Using Eq. \ref{eq:linear_response}, one can show that the response is of the form
\begin{align} \label{eq:response}
&K^{\mu\nu}_{n,n'}(q,q') = g^2T \sum\limits_{m}\int \frac{d^dp}{(2\pi)^d} \Gamma^{\mu\nu}_{m,-n,-n'}(p,-q,-q') \langle\phi^\dagger_{m-n-n'}(p-q-q')\phi_m(p)\rangle  \nonumber \\
&  -g^2T^2\sum\limits_{m,m'} \int \frac{d^dp}{(2\pi)^d}\frac{d^dp'}{(2\pi)^d}\Gamma^\nu_{m,-n'}(p,-q')\Gamma^{\mu}_{m',-n}(p',-q) \langle\phi^{\dagger}_{m-n'}(p-q')\phi_m(p)\phi^\dagger_{m'-n}(p'-q)\phi_{m'}(p')\rangle.
\end{align}
Expanding the four-point correlation function with Wick's theorem and substituting
\beq
\langle \phi^\dagger_m(p)\phi_n(p') \rangle = \frac{(2\pi)^d}{T} \delta(p-p')\delta_{m,n}G_m(p) \label{eq:two_point}
\eeq
into Eq. \ref{eq:response}, one finds
\begin{align}
&K^{\mu\nu}_{n}(q) = g^2T\sum\limits_{m}\int \frac{d^dp}{(2\pi)^d} \Gamma^{\mu\nu}_{m,-n,n}(p,-q,q)G_m(p) \nonumber \\
&  -g^2T\sum\limits_{m} \int\frac{d^dp}{(2\pi)^d}\Gamma^\nu_{m,n}(p,q)\Gamma^{\mu}_{m+n,-n}(p+q,-q)G_m(p)G_{m+n}(p+q).
 \label{eq:response_general}
\end{align}
The first and the second terms of Eq. \ref{eq:response_general} are known as the diagmagnetic and the paramagnetic terms, respectively. Summing over the Matubara frequency with the standard contour integration technique (see Appendix \ref{app:summation_matsubara}), one finds the optical conductivity of the scalar unparticle (Eq. \ref{eq:final_conductivity}) is
\beq
\sigma(i\omega_n) = (\frac{d+1}{2} - d_U) \sigma_{0}(i\omega_n) \label{eq:con_unparticle_eff_action}
\eeq
where $\sigma_{0}$ is the conductivity of a free massless scalar field (Eq. \ref{eq:massive_scalar_conductivity} with m = 0). In $d=3$, the prefactor is $2-d_U$.  Liao\cite{Liao2008} showed that the polarization is the same prefactor, $2-d$, times the polarization for the scalar particle field, thereby corroborating the lengthy calculation performed here.  Note that, when  $d_U < \frac{d+1}{2}$, the real part of the conductivity becomes negative, which is not physical. Thus, the allowed value for $d_U$ in a Gaussian model is $\frac{d-1}{2} < d_U < \frac{d+1}{2}$. From the form of $\sigma_0$ in Eqs. \ref{eq:massive_scalar_conductivity_real} and \ref{eq:massive_scalar_conductivity_imag}, we see that $\sigma_0$ has no dependence on $d_U$. Therefore, the optical conductivity from the unparticle effective action does not have a fractional power law.  In the context of the polarization for unparticles.

At work here is the linear superposition that links the unparticle to the particle field, inherent in a Gaussian action for unparticles.  Such a superposition implies that the statistics of the unparticle cannot deviate from that of the particles.  Also at work here is the fact that vertex corrections within a Gaussian action kill the power law at the bare one-loop level.  This can be seen as follows.  The scaling dimension of the unparticle propagator is [L]$^{d+1-2d_U}$ where [L] denotes a scaling in units of length. We know that the vertex couplings must satisfy the two Ward-Takahashi identities. So $\Gamma^{\mu}$ and $\Gamma^{\mu\nu}$ have a scaling dimension of [L]$^{2d_U-d}$ and [L]$^{2d_U-d+1}$, respectively. As a result, the scaling dimension of the response function (Eq. \ref{eq:response_general}) is [L]$^{1-d}$ and that of the conductivity (Eq. \ref{eq:sigma_K}) is [L]$^{2-d}$. In other words, the anomalous length scale in the unparticle propagator, namely, anything that has a scale ~[L]$^{d_U}$ or its integer power, is cancelled by the vertices.  This applies equally to the Luttinger liquid calculation\cite{Anderson1997}.

\section{Optical Conductivity from the Continuous Mass Formalism} \label{sec:optical_sum}

In this section, we apply the idea of the continuous mass formalism discussed in section \ref{sec:mass_sum}. We consider a model with large number of flavors of free fields or free particles. We sum over the conductivity of each flavor to obtain the total conductivity of this ensemble of free fields. In the continuous mass formulation of unparticles we explained in section \ref{sec:mass_sum}, there is no cutoff in the mass integration. Nonetheless as we will see, we take $M$ to be a mass cutoff in the calculations below. 

\subsection{Scalar Field} \label{sec:optical_sum_scalar}
We consider a model system with a large number of free complex scalar fields,
\beq
S = \sum\limits_{i=1}^N\int d\tau \int d^dx (|D_{\mu}\phi_i|^2+m_i^2|\phi_i|^2), 
\eeq
where the covariant derivative is defined in the usual way as $D_\mu = \partial_\mu - ie_iA_\mu$ and the summation is over all flavors. The mass and charge of the field of flavor $i$ are given by $m_i$ and $e_i$, respectively. Let the density of flavors, $\rho(m)$, to be a function of mass in the mass range from $0$ to $M$. The number of fields in the mass range $m$ to $m+dm$ is $\rho(m)dm$. Furthermore, we let the charge to be a function of mass, $e(m)$. The forms of $\rho(m)$ and $e(m)$ are chosen in the same way as in Ref. \cite{Karch2015}:
\beq
\rho(m) &=& \frac{m^{a-1}}{M^a} \rho_0 \\
e(m) &=& \frac{m^b}{M^b} e_0
\eeq
Here $a$ and $b$ are the power law scaling of the density of flavors and charges. As pointed out previously \cite{Karch2015}, $a$ and $b$ are related to the hyperscaling violation exponent and the anomalous dimension for the current, respectively.

\begin{figure}
        \centering
        \subfigure[$\ $ Real part
        \label{fig:scalar_con1_sum}]{\includegraphics[scale=0.35]{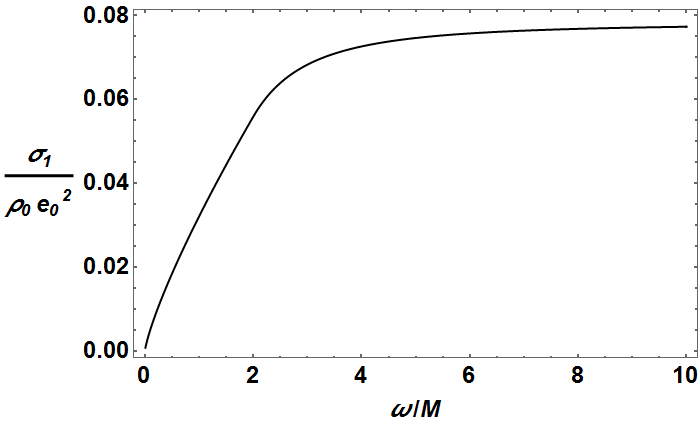}}
        \subfigure[$\ $ Imaginary part \label{fig:scalar_con2_sum}]{\includegraphics[scale=0.35]{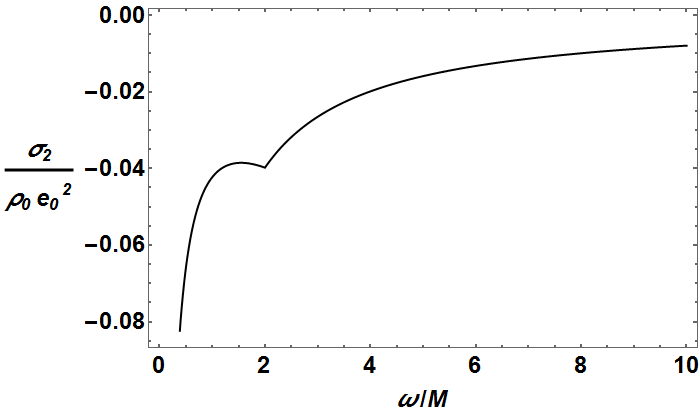}}
	\caption{Real part and imaginary part of the total conductivity in 2D at zero temperature for the case a = 0.2 and b = 0.3}
\end{figure}

From Eq. \ref{eq:massive_scalar_conductivity} in appendix \ref{app:scalar}, the optical conductivity of flavor $i$ has the form
\beq
\sigma^i(\omega) = e^2_if(\omega,m_i,T). \label{eq:sigma_i}
\eeq
The total conductivity is a summation of conductivity of all flavors,
\beq
\sigma(\omega) &=& \sum\limits_{i}\sigma^i(\omega) \nonumber \\
&=& \int\limits_{0}^{M} dm \ \rho(m)e^2(m)f(\omega,m,T). \label{eq:sigma_summation}
\eeq
We replace the summation by the integral over mass and the density function $\rho(m)$. Rewriting the conductivity of a massive free scalar field (Eqs. \ref{eq:massive_scalar_conductivity_real} and \ref{eq:massive_scalar_conductivity_imag}) in the form of Eq. \ref{eq:sigma_i} and scaling out the frequency, one obtains
\beq
\sigma^i_1(\omega) = \mathrm{Re} \sigma^i(\omega) &=& \pi \rho^i_D \delta(\omega) + e^2_i  \omega^{d-2}f_1(\frac{m_i}{\omega},\frac{T}{\omega}) \\
\sigma^i_2(\omega) = \mathrm{Im} \sigma^i(\omega) &=& \frac{\rho^i_D}{\omega} + e^2_i \omega^{d-2} f_2(\frac{m_i}{\omega},\frac{T}{\omega}).
\eeq
The Drude weight $\rho^i_D$ is given by
\beq
\rho^i_D = \frac{2e^2_i \beta S_d}{d(2\pi)^d} \int\limits_{0}^{\infty} dp \ p^{d+1} \frac{N(\varepsilon_p)(N(\varepsilon_p)+1)}{\varepsilon_p^2}
\eeq
where $N(E) = \frac{1}{e^{\beta E}-1}$ is the Bose-Eisnstein distribution, $S_{d} = \frac{2\pi^{\frac{d}{2}}}{\Gamma(\frac{d}{2})}$ is a surface area of a unit (d-1)-sphere, and $\varepsilon_p = \sqrt{p^2+m_i^2}$. The functions $f_1$ and $f_2$ are given by
\beq
f_1\bigg(\frac{m}{\omega},\frac{T}{\omega}\bigg) &=& \frac{\pi S_d}{d 2^{d} (2\pi)^d} \theta\bigg( \bigg(\frac{\omega}{m} \bigg)^2-4 \bigg)\bigg(1-\frac{4m^2}{\omega^2}\bigg)^{\frac{d}{2}} \coth\bigg(\frac{\beta\omega}{2}\bigg) \nonumber \\
f_2\bigg(\frac{m}{\omega},\frac{T}{\omega}\bigg) &=& \frac{S_d}{d(2\pi)^d}\int\limits_{0}^{\infty} dy \frac{y^{d+1}}{(y^2+(\frac{m}{\omega})^2)^{\frac{3}{2}}}P\bigg( \frac{1}{1-4(y^2+(\frac{m}{\omega})^2)}\bigg)\coth{\bigg(\frac{\beta\omega\sqrt{y^2+(\frac{m}{\omega})^2}}{2}\bigg)}, \nonumber
\eeq
where $\theta(x)$ is the Heaviside function and $P$ denotes the Cauchy principal integral. We consider a special case of $d = 2$ and $T = 0$ (Eqs. \ref{eq:scalar_conductivity_d2T0_real} and \ref{eq:scalar_conductivity_d2T0_imag}). In this case, the Drude weight $\rho^i_D = 0$ and the integral in $f_2$ can be done analytically. One finds
\beq
\sigma^i_1(\omega) &=& e^2_i f_1(\frac{m_i}{\omega}) \\
\sigma^i_2(\omega) &=& e^2_i f_2(\frac{m_i}{\omega})
\eeq
and
\beq
f_1\bigg(\frac{m}{\omega}\bigg) &=& \frac{1}{16}\theta\bigg(\bigg(\frac{\omega}{m}\bigg)^2 -4\bigg) \bigg(1- 4\bigg(\frac{m}{\omega}\bigg)^2 \bigg) \nonumber \\
f_2\bigg(\frac{m}{\omega}\bigg) &=& \frac{1}{16\pi}\bigg(-4\frac{m}{\omega} + (1-4(\frac{m}{\omega})^2)\ln \bigg|\frac{1-2\frac{m}{\omega}}{1+2\frac{m}{\omega}}\bigg|\bigg). \nonumber
\eeq
From Eq. \ref{eq:sigma_summation}, the real part of conductivity is
\beq
\sigma_1(\omega) = \mathrm{Re} \sigma(\omega) &=& \int\limits_{0}^{M} dm \ \rho(m) e^2(m)  f_1(\frac{m}{\omega}) \nonumber \\ 
&=& 
\begin{cases}
\frac{1 }{2^{a+2b+3}(a+2b)(a+2b+2)}\frac{\rho_0e^2_0\omega^{a+2b}}{M^{a+2b}}  \ \ \  \text{if } \omega < 2M \\
\frac{\rho_0e^2_0}{4}(\frac{1}{4(a+2b)}- \frac{1}{(a+2b+2)}(\frac{M}{\omega})^{2}) \ \ \  \text{if } \omega > 2M 
\end{cases}
\eeq
where $a+2b$ must be greater than $0$ in order for the integral to converge. When $\omega/M < 2$, $\sigma_1$ has a fractional power law of $\omega^{a+2b}$. However, when $\omega/M > 2$, $\sigma_1$ is approaching an asymptotic value with a power law of $\omega^{-2}$. A plot of $\sigma_1$ in the two-dimensional case at zero temperature is shown in Fig. \ref{fig:scalar_con1_sum}.  

The imaginary part of conductivity can be evaluated approximately as
\beq
\sigma_2(\omega) = \mathrm{Im} \sigma(\omega) &=& \int\limits_{0}^{M} dm \ \rho(m) e^2(m)  f_2(\frac{m}{\omega}) \\
&\approx&  
\begin{cases}
-\frac{3\rho_0e^2_0}{16\pi(a+2b+1)}\frac{M}{\omega} - \frac{5\rho_0e^2_0}{2^{a+2b+5}\pi(a+2b+1)}(\frac{\omega}{M})^{a+2b}
  \ \ \  \text{if } \omega < 2M  \nonumber \\
-\frac{\rho_0e^2_0}{2\pi(a+2b+1)}\frac{M}{\omega} \ \ \  \text{if } \omega > 2M, 
\end{cases}
\eeq
where $a+2b$ must be greater than $0$ in order for the integral to converge. Unlike $\sigma_1$, when $\omega/M < 2$, $\sigma_2$ depends on the terms $\sim \omega^{a+2b}$ and $\sim\omega^{-1}$. When $\omega/M > 2$, $\sigma_2$ is proportional to $\omega^{-1}$. A plot of the approximated $\sigma_2$ in the two-dimensional case at zero temperature is shown in Fig. \ref{fig:scalar_con2_sum}. The real part of scalar conductivity in two dimension at finite temperature is
\beq
\sigma_1(\omega) =  \begin{cases}
\pi \rho_D \sigma(\omega) + \frac{1}{2^{a+2b+3}(a+2b)(a+2b+2)}\frac{\rho_0e^2_0\omega^{a+2b}}{M^{a+2b}}  \coth(\frac{\beta \omega}{2}) \ \ \  \text{if } \omega < 2M \\
\frac{\rho_0e^2_0}{4}(\frac{1}{4(a+2b)}- \frac{1}{(a+2b+2)}(\frac{M}{\omega})^{2})\coth(\frac{\beta \omega}{2}) \ \ \ \text{if } \omega > 2M,
\end{cases}
\eeq
where $\rho_D$ is the total Drude weight and is given by
\beq
\rho_D = \frac{\rho_0e^2_0 \beta}{2\pi M^{a+2b}} \int\limits_{0}^{M} dm\int\limits_{0}^{\infty} dp \ m^{a+2b-1}p^{3} \frac{N(\varepsilon_p)(N(\varepsilon_p)+1)}{\varepsilon_p^2}. \nonumber 
\eeq
The imaginary part of the conductivity at finite temperature can be obtained from the Kramers-Kronig relationship. At finite temperature, $\sigma_1$ is modified by a bosonic factor, $\coth(\frac{\beta \omega}{2})$. Consequently, for $T\ll M$, $\sigma_1$ still has a power law of $\omega^{a+2b}$ when $\omega<2M$. When $T$ is about the same order of magnitude as $M$ or larger, the power law in $\sigma_1$ no longer exists. Hence, although we have obtained a fractional power law in the case of a scalar field, it is a power law with a positive power not negative power ($a+2b > 0$) as required to explain the experiments.

\subsection{Drude Conductivity} \label{sec:optical_sum_drude}
Motivated by the experiments\cite{Marel2003}, in which the real and imaginary parts have a negative power law, we apply the continuous mass formalism to the Drude form of the conductivity.  The system consists of large number of flavors of free non-relativistic fermions in a disordered potential with no interaction among different flavors of fermions. The conductivity of one flavor is simply a Drude conductivity \cite{Altland2010},
\beq
\sigma^i(\omega) = \frac{n_i e^2_i \tau_i}{m_i} \frac{1}{1-i\omega \tau_i}
\eeq
where $m_i$, $e_i$, $n_i$, and $\tau_i$ are mass, charge, density, and relaxation time of fermion of flavor $i$. As in the case of a scalar field, we take the density of flavors, $\rho(m)$, charge $e(m)$, and relaxation time, $\tau(m)$, all to be functions of mass. The mass ranges from $0$ to $M$. The charge density in the mass range of $m$ to $m+dm$ is $dn(m) = \rho(m)dm$. So the contributions to the conductivity from the flavors in the range $m$ to $m+dm$ is
\beq
d\sigma = \frac{\rho(m)e^2(m)\tau(m)}{m}\frac{1}{1-i\omega\tau(m)}dm.
\eeq
The conductivity of all flavors is
\beq
\sigma(\omega) = \int\limits_{0}^{M}\frac{\rho(m)e^2(m)\tau(m)}{m}\frac{1}{1-i\omega\tau(m)} dm.
\eeq
We choose the form of the charge, density, and relaxation time in the same way in the scalar field case, namely,
\beq
\rho(m) &=& \rho_0\frac{m^{a-1}}{M^a} \\
e(m) &=& e_0\frac{m^b}{M^b} \\
\tau(m) &=& \tau_0\frac{m^c}{M^c}.
\eeq
The conductivity becomes
\beq
\sigma(\omega) = \frac{\rho_0e^2_0\tau_0}{M^{a+2b+c}}\int\limits_{0}^{M}dm \frac{m^{a+2b+c-2}}{1-i\omega \tau_0\frac{m^c}{M^c}}.
\eeq
Changing the integration variable to $x = \omega\tau_0\frac{m^c}{M^c}$, one obtains
\beq
\sigma(\omega) = \frac{\rho_0e^2_0}{cM}\frac{1}{\omega(\omega\tau_0)^{\frac{a+2b-1}{c}}}\int\limits_{0}^{\omega\tau_0}dx \ \frac{x^{\frac{a+2b-1}{c}}}{1-ix}.
\eeq
Define $\alpha\equiv\frac{a+2b-1}{c}+1$. When $\alpha<0$, the integral does not converge. This means it is {\it not} possible to obtain a positive power law in the optical conductivity from this model. When $0<\alpha<1$, the integral has a simple form in the high freqeuency limit $\omega \tau_0 \gg 1$:
\beq
\sigma(\omega) = \frac{\rho_0e^2_0}{cM\tau_0^{\alpha - 1}}\frac{\pi e^{\frac{i\alpha\pi}{2}}}{\sin \pi\alpha}\omega^{-\alpha}.
\eeq
Since $\mathrm{Re}\sigma(\omega)$ falls off slower than $\frac{1}{\omega}$, the total spectral weight, $\int\limits_{0}^{\infty}\mathrm{Re}\sigma(\omega)d\omega$ is infinite. So we need to define a high frequency cutoff, $\omega_C$, to have a finite spectral weight.\footnote{{Even though the spectral weight of one flavor is finite,\begin{equation*} \int\limits_{0}^{\infty}\frac{n_i e^2_i \tau_i}{m_i} \frac{1}{1+\omega^2 \tau_i^2} d\omega = \frac{\pi n_i e^2_i }{2 m_i}, \end{equation*} the integral of the total spectral weight, $\int\limits_{0}^{M} dm \frac{\pi \rho_0 e_0^2 }{2 }m^{a+2b-2} $, diverges when $0<\alpha<1$ and $c$ is a positive number. Here $c$ must be positive, otherwise the real part of conductivity would be negative. We still need to define a cutoff, $\omega_C$, in order for the total spectral weight calculated this way to be finite.}} Here we see that by choosing the value of $a$, $b$, and $c$ such that $0<\alpha<1$, the conductivity can have a fractional power law with a power between -1 and 0 and a phase of $\frac{\alpha \pi}{2}$. 

As a concrete example, let us try to reproduce the power law of $-\frac{2}{3}$ seen in the optimally doped cuprates above the superconducting dome \cite{Marel2003}. We choose $\frac{a+2b-1}{c} = -\frac{1}{3}$ with $c = 1$, so $\alpha = \frac{2}{3}$. The conductivity is
\beq
\sigma(\omega) &=& \frac{\rho_0e^2_0\tau^{\frac{1}{3}}_0}{M}\frac{1}{\omega^{\frac{2}{3}}}\int\limits_{0}^{\omega\tau_0} dx \ \frac{x^{-\frac{1}{3}}}{1-ix} \nonumber \\
&=& \frac{\rho_0e^2_0\tau^{\frac{1}{3}}_0}{M}\frac{1}{\omega^{\frac{2}{3}}} \bigg(\ln(1+i(\tau_0\omega)^{\frac{1}{3}})+e^{\frac{2i\pi}{3}}\ln(1-e^{\frac{i\pi}{6}}(\tau_0\omega)^{\frac{1}{3}}) - e^{\frac{i\pi}{3}}\ln(1-e^{\frac{5i\pi}{6}}(\tau_0\omega)^{\frac{1}{3}})\bigg) ,
\eeq 
where the branch cut of the logarithm is chosen to be along the negative real axis. The cutoff is set at some high frequency $\omega_C$. Taking the limit of $\tau_0\omega \rightarrow \infty$, we obtain
\beq
\sigma(\omega) = \frac{1}{3}(\sqrt{3}+3i)\pi\frac{\rho_0 e^2_0 \tau^{\frac{1}{3}}_0}{M\omega^\frac{2}{3}}.
\eeq
We see that the power-law exponent is $-\frac{2}{3}$ with a phase of $60^\circ$ as is seen experimentally.  While there are ways of generating fractional exponents\cite{Hartnoll2011,Chubukov2014} in the optical conductivity, none yield the exponent of $-2/3$.  The recent claim that gravitational crystals\cite{santos2} produce the $\omega^{-2/3}$ conductivity of the cuprates was not borne out by subsequent calculations\cite{Donos2014,Rangamani2015,Langley}.   Consequently, the results obtained here are the first to yield the $-2/3$ exponent from a strong coupling formulation.  Our result is general, following strictly from the continuous mass formalism and the Drude form for the conductivity.  Hence, what is needed of a microscopic theory is one in which the optical conductivity at each frequency receives contributions from all energy scales.  That is, incoherence is at the heart of the power law optical conductivity in strongly correlated electron systems.

When $\alpha > 1$, the conductivity still has a power law of $\omega^{-\alpha}$ in the real part and there is no need for the high frequency cutoff. For example, we choose $\frac{a+2b-1}{c} = \frac{1}{5}$ with $c = 1$, so $\alpha = \frac{6}{5}$. The conductivity in the limit of high frequency is
\beq
\sigma(\omega) = \frac{\rho_0e^2_0}{M}(\frac{5i}{\omega} + \frac{1.65163 - 5.0832i}{\tau_0^{\frac{1}{5}}\omega^\frac{6}{5}}).
\eeq
The real part of the conductivity has a power of $-\frac{6}{5}$, whereas the imaginary part has an extra term $\sim\omega^{-1}$. 

\section{Discussion and Implications}

The result that there is no fractional power law from a Gaussian form of unparticle action is not a surprise as follows from the scaling form of the vertex.   On the other hand, other response functions that do not have this dimensional cancellation from vertices should exhibit a non-integer power that depends on $d_U$. One example of this kind of response function is the density-density correlation
 $\langle\phi^\dagger(x)\phi(x)\phi^\dagger(y)\phi(y)\rangle$. 

From this reasoning, one might think it is possible to obtain a power law if we know a way to embed an anomalous dimension for the current into the effective action. The present approach we used to gauge the action using Wilson lines does not allow one to do so. If we can find another way to write down a gauged effective action with the anomalous dimension for the current included, one should be able to get a power law that depends on the anomalous dimension. However, from the result of the mass summation over the two-dimensional scalar field at T = 0, we find that the exponent is $a+2b$, where $a$ is related to a hyperscaling violation exponent (or $d_U$) and $b$ is related to anomalous dimension for the current. Clearly, even when $b=0$, it is still possible to have a power law. This signifies that one can nevertheless obtain a power law in the optical conductivity without the inclusion of the anomalous dimension for the current.  

The fact that the conductivity of the Gaussian unparticle is the same as that of massless scalar particle up to a constant multiplication factor is interesting. The technical reason of this result is due to an exact cancellation of discontinuities across branch cuts of diamagnetic and paramagnetic terms. The response functions that do not have this kind of cancellation, such as density-density correlations, should acquire non-trivial dependence on the anomalous dimension, $d_U$, unlike the case of the optical conductivity.

We want to mention several points about the scalar field summation results. First, in d = 2 and T = 0 case, $\sigma_1$ has a positive power law. This is due to the fact that we sum over a collisionless contribution to the conductivity (see \cite{Damle1997} for an explanation of each part of the conductivity). Should we add a momentum relaxation process, for example, dissipation and interactions, and then sum over this contribution to the conductivity we would get a  negative power law. Phenomenologically, this part of the conductivity should have a Drude form. We perform a similar calculation on the Drude conductivity in the section \ref{sec:optical_sum_drude}. We find that, indeed, the conductivity in that case has a negative fractional power law. The second point is that $\sigma_1$ in the case of d = 2 and T = 0 has a fractional power law, whereas $\sigma_2$, in addition to a term $\sim \omega^{a+2b}$, has a term that goes like $\sim \omega^{-1}$. At low frequency, $\sigma_2$ is dominated by $\omega^{-1}$ behavior. At higher frequency, the power law of $\omega^{a+2b}$ could appear in the frequency range $M<\omega<2M$, if the exponent $a+2b$ is large enough. This result suggests that in a system in which $\sigma_1$ has a positive fractional power law, $\sigma_2$ may change from having an integer power law at low frequency to a fractional power law at higher frequency. In such cases, the power law disappears at finite temperature due to a bosonic temperature factor, $\coth(\frac{\beta \omega}{2})$. As long as the temperature is kept to be much lower than $M$, there is still a power law in $\sigma_1$. When the temperature is raised to be about the same order of magnitude as $M$ or higher, the power law disappears.

The key result of this paper is that continuous mass formalism coupled with the Drude conductivity yields a power law with exponents between -1 and 0 which are in agreement with the exponents seen in the cuprates.  The key point of the continuous mass formalism is that summation over mass is summation over energy.  Hence, the power law found here arises from the incoherent part of the spectrum, that is the part that has no particle content, hence the description in terms of unparticles.    As pointed out in section \ref{sec:optical_sum_drude}, one needs to set a high frequency cutoff in order to have a finite spectral weight. While the continuous mass formalism works well as a phenomenological tool to explain an appearance of power laws,  it cannot produce other features that require microscopic details. Our results suggest that a microscocpic theory that can explain fractional power laws should be decomposable into a large number of free fields or free particles with varying masses and possibly varying charges. Still, one should keep in mind that there could be other models that predict a power law. We should also note the result when the exponents are less than -1. $\sigma_1$ has a fractional power law $\omega^{-\alpha}$  at high frequency, but $\sigma_2$ behaves like $A\omega^{-\alpha} + B\omega^{-1}$ where A and B are real constant.

To conclude, we studied the optical conductivity in two systems: a system of unparticles with a Gaussian form for the action and a system with a large number of flavors of free fields. We found that the conductivity of the Gaussian unparticle does not have a fractional power law that depends on $d_U$. However, using the continuous mass formalism on the system with a large number of free fields, we obtained a positive fractional power law in the case of free scalar field and negative fractional power law in the case of Drude conductivity.  That the combination of exponents here can be constrained by other transport measurements (Hall angle and DC conductivity) on the cuprates will be the subject of a forthcoming publication.  The latter result points to two possibilities for a microscopic model: either an anomalous dimension is present which indicates some kind of multi-band system or the anomalous dimension arises from UV-IR mixing.  This has been pointed out extensively in the Hubbard model\cite{Phillips2013}.  Hence, perhaps it might be able to construct the power-law conductivity from a theory in which the upper Hubbard is integrated out. Ref. \cite{tpc2008} relied on the non-crossing approximation and hence any effort forward must be based on new techniques that can incorporate the vertex corrections when a particle-picture breaks down.   \\

\textbf{Acknowlegements}
We thank Jimmy Hutasoit, Victor Chua, Brandon Langley, Tony Hegg, Zhidong Leong, and Garrett Vanacore for sustained commentary throughout the completion of this work and NSF DMR-1461952 for partial funding of this project. KL is supported by the Department of Physics at the University of Illinois and a scholarship from the Ministry of Science and Technology, Royal Thai Government.  PP thanks the Guggenheim Foundation for a 2015-2016 Fellowship.

\appendix
\section{Optical Conductivity of a Free Massive Scalar Field} \label{app:scalar}
The optical conductivity of the free massive scalar field has been studied by many authors (for exmaple, see Refs. \cite{Otterlo1993,Damle1997}). Since the scalar field conductivity is used in section \ref{sec:optical_sum} and appendix \ref{app:summation_matsubara}, we briefly review the basic formulation and the main results in this appendix. The optical conductivity can be calculated from Eq. \ref{eq:response_general}. The propagator of the massive scalar field in Matsubara space is
\beq
G_n(k) = \frac{1}{\omega_n^2 + \varepsilon^2_k} \label{eq:scalar_propagator}
\eeq
where the bosonic Matsubara frequency $\omega_n = \frac{2\pi n}{\beta}$ (n being an integer), k is a d-dimensional momentum, $\varepsilon_k = \sqrt{k^2 + m^2}$, and $m$ is the mass of the scalar field. The vertices in this case are
\beq
\Gamma^\mu_{m,n}(p,q) = (2p+q)^{\mu} \label{eq:scalar_v1} \\ 
\Gamma^{\mu\nu}_{m,n_1,n_2}(p,q_1,q_2) = 2\delta^{\mu\nu}. \label{eq:scalar_v2}
\eeq
For the expression $p^\mu$, $p^0$ is a Matsubara frequency and $p^i$, for $i$ = 1 to d, is an ith component of d-dimensional momentum. Substituting Eqs. \ref{eq:scalar_propagator}, \ref{eq:scalar_v1}, and \ref{eq:scalar_v2} into Eq. \ref{eq:response_general}, one finds the longitudinal conductivity (the $ii$ component with $i\neq0$) in the limit $q\rightarrow 0$ to be
\beq
\sigma(i\omega_n) = \frac{2g^2}{\omega_n} T\sum\limits_{m} \int \frac{d^dp}{(2\pi)^d} \bigg( \frac{1}{\omega_m^2 + \varepsilon^2_p} - \frac{2p^2_i}{(\omega^2_m + \varepsilon^2_p)((\omega_m+\omega_n)^2 + \varepsilon^2_p)}\bigg).
\eeq
Inserting $ 1 = \frac{dp_i}{dp_i}$ into the diamagnetic term and integrating by parts, one finds
\beq
\sigma(i\omega_n) = \frac{4g^2}{\omega_n} T\sum\limits_{m} \int \frac{d^dp}{(2\pi)^d} p^2_i\bigg( \frac{1}{(\omega_m^2 + \varepsilon^2_p)^2} - \frac{1}{(\omega^2_m + \varepsilon^2_p)((\omega_m+\omega_n)^2 + \varepsilon^2_p)}\bigg).
\eeq
The boundary term $\sim T\sum\limits_{m} \int d^{d-1}p \ \frac{p_i}{\omega_m^2 + \varepsilon^2_p}\big|^\infty_{p_i = -\infty} $ vanishes in the dimensional or lattice regularization \cite{Damle1997}. We can also see this clearly in the case of a lattice system with periodic boundary conditions. In this case, $p_i$ is replaced by $\sin p_i$ and $\pm\infty$ becomes the momenta at the edge of the Brillouin zone $= \pm\pi$. As a result, the boundary term vanishes (see appendix B of Ref. \cite{Cha1991}). Had we not done the integration by parts, the optical conductivity would have a zero-frequency delta function peak at zero temperature. We implicitly set the chemical potential of the scalar field to  0. According to the Bose-Eisenstein distribtution, there are no particles or antiparticles present at zero temperature. This means that the dc conductivity should vanish and thus the zero-frequency delta function is not physical. At finite temperature, the mixture of particles and antiparticles is excited into the system and so the zero-frequency delta function should appear. 

\begin{figure}
        \centering
        \subfigure[$\ $ Diamagnetic term
        \label{fig:scalar_dia}]{\includegraphics[scale=0.5]{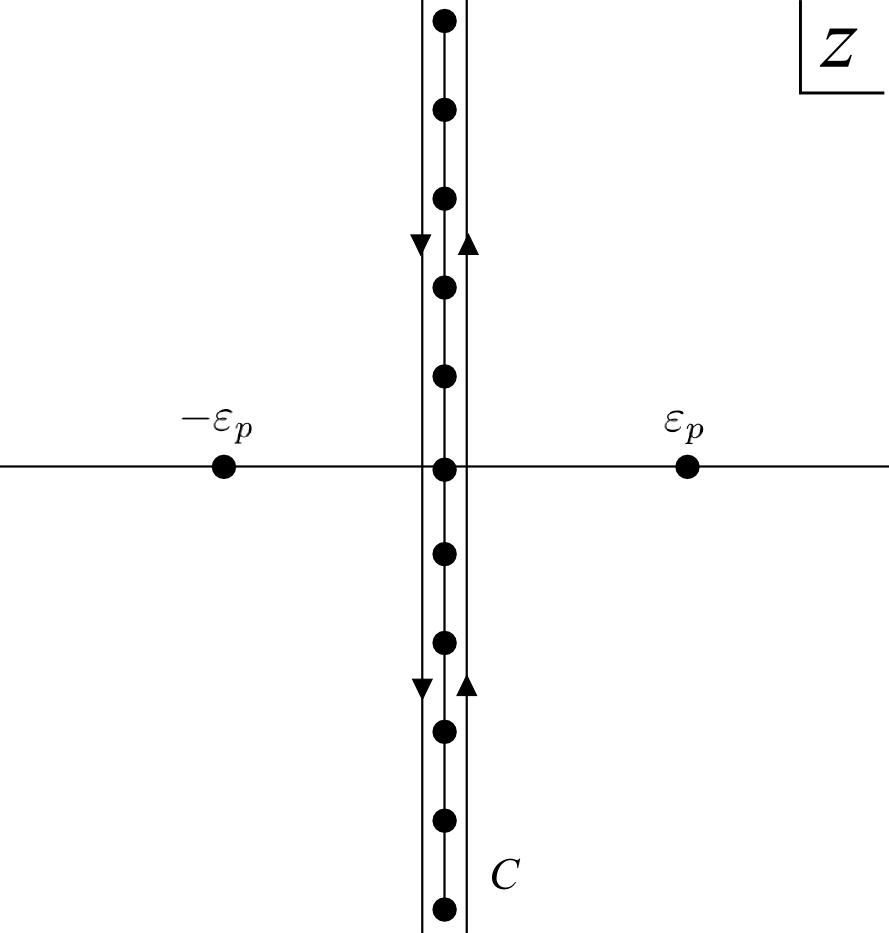}}
        \subfigure[$\ $ Paramagnetic term \label{fig:scalar_para}]{\includegraphics[scale=0.5]{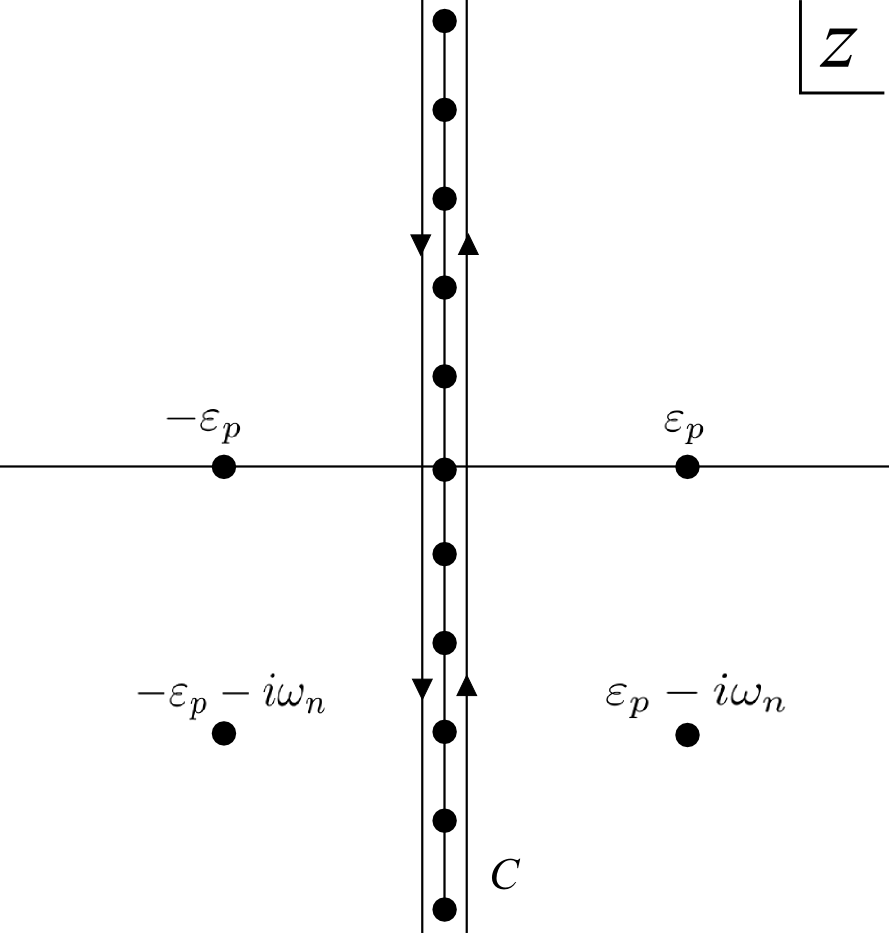}}
	\caption{The contours used in the Matsubara summation of the diamagnetic and paramagnetic terms. }
\end{figure}

\begin{figure}
        \centering
        \subfigure[$ \ $ Real part \label{fig:2d_real}]{\includegraphics[scale=0.3]{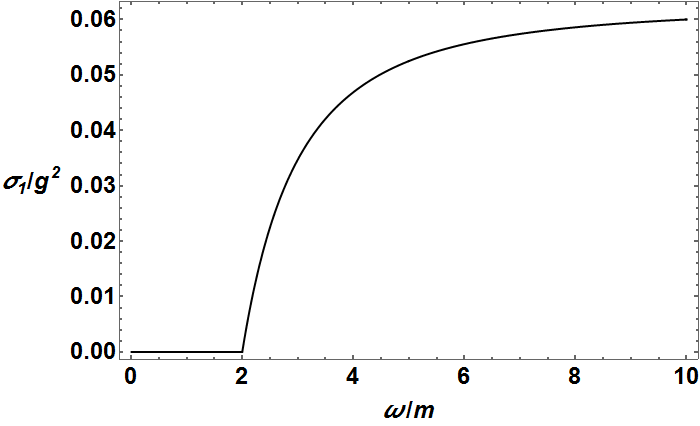}}
        \subfigure[$ \ $ Imaginary part \label{fig:2d_imag}]{\includegraphics[scale=0.3]{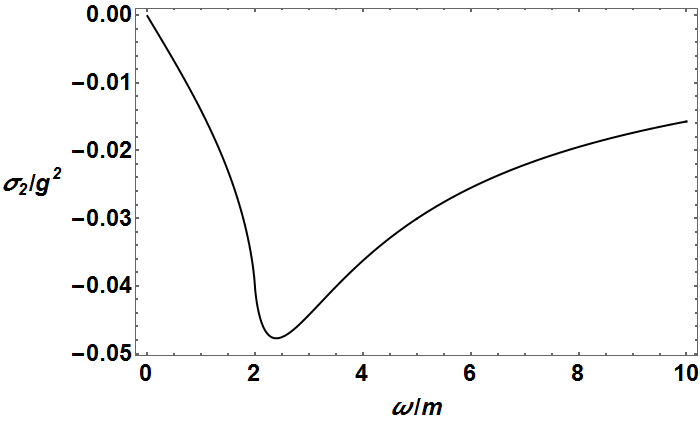}}
        \caption{Conductivity of massive scalar field in d = 2 at zero temperature. \label{fig:2d}}
\end{figure}

We perform Matsubara a summation explicitly using contour integration. Let us first consider the summation on the diamagnetic term. Converting the summation into a contour integral with the contour shown in Fig. \ref{fig:scalar_dia}, we have
\beq
 T\sum\limits_{m}\frac{1}{(\omega_m^2 + \varepsilon^2_p)^2} = \int_C \frac{dz}{2\pi i} \frac{g(z)}{(z^2-\varepsilon_p^2)^2},
\eeq
with $g(x) = \frac{1}{2}\coth(\frac{\beta x}{2})$.  Upon deforming the contour $C$ to infinity, the contour integral turns into a summation of the residues at $\pm \varepsilon_p$,
\beq
T\sum\limits_{m}\frac{1}{(\omega_m^2+\varepsilon_p^2)^2} &=& -\sum\limits_{z_i=\pm\varepsilon_p}Res_{z_i}\bigg[ \frac{g(z)}{(z^2-\varepsilon_p^2)^2} \bigg] \nonumber \\
&=& \frac{\beta}{2\varepsilon_p^2}N(\varepsilon_p)(N(\varepsilon_p)+1) + \frac{1}{2\varepsilon_p^3}g(\varepsilon_p),\nonumber
\eeq
where $N(x) = \frac{1}{e^{\beta x}-1}$ is the Bose-Einstein distribution.
The similar computation, with the contour shown in Fig. \ref{fig:scalar_para}, can be done on the paramagnetic term to show that
\beq
T\sum\limits_{m}\frac{1}{(\omega_m^2+\varepsilon_p^2)((\omega_m+\omega_n)^2+\varepsilon_p^2)} =  -\frac{g(\varepsilon_p)}{i\omega_n\varepsilon_p}\bigg(\frac{1}{i\omega_n+2\varepsilon_p} - \frac{1}{-i\omega_n + 2\varepsilon_p} \bigg).
\eeq
Combining the results for the paramagnetic and the diamagnetic terms, one finds the conductivity in  Matsubara space to be
\beq
\sigma(i\omega_n) = 2g^2\int \frac{d^dp}{(2\pi)^d} p^2_i \Biggl\{  \frac{\beta}{\omega_n p^2}N(\varepsilon_p)(N(\varepsilon_p)+1) + \frac{ig(\varepsilon_p)}{2\varepsilon_p^3}\bigg(\frac{1}{2\varepsilon_p+i\omega_n}-\frac{1}{2\varepsilon_p-i\omega_n} \bigg) \Biggl\}. \label{eq:massive_scalar_conductivity}
\eeq
Pefroming analytic continuation $i\omega_n \rightarrow \omega+i0^+$, the conductivity turns out to be
\beq
\sigma_1(\omega) = \mathrm{Re} \sigma(\omega) &=& \pi\rho_D\delta(\omega) + \frac{g^2\pi S_d}{d2^{d}(2\pi)^d} \theta(\omega^2 - (2m)^2)\omega^{d-2}(1-\frac{(2m)^2}{\omega^2})^{d/2}\coth(\frac{\beta\omega}{2}) \label{eq:massive_scalar_conductivity_real} \\
\sigma_2(\omega) = \mathrm{Im} \sigma(\omega) &=& \frac{\rho_D}{\omega} + \frac{g^2S_d}{d(2\pi)^d}\int\limits_{0}^{\infty}dp \frac{p^{d+1}}{\varepsilon^3_p}P(\frac{\omega}{\omega^2-4\varepsilon^2_p})\coth(\frac{\beta\varepsilon_p}{2}), \label{eq:massive_scalar_conductivity_imag}
\eeq
where $\theta(x)$ is the Heaviside function, $P$ denotes the Cauchy principal integral, $S_{d}$ = $\frac{2\pi^{\frac{d}{2}}}{\Gamma(\frac{d}{2})}$ the surface area of a unit (d-1)-sphere and the Drude weight $\rho_D$ is given by
\beq
\rho_D = \frac{2g^2\beta S_d}{d(2\pi)^d}\int\limits_{0}^{\infty}dp \  p^{d+1}\frac{N(\varepsilon_p)(N(\varepsilon_p)+1)}{\varepsilon^2_p}.
\eeq
Note that at T = 0, $\rho_D$ = 0. So the zero-frequency delta function vanishes at T = 0 not at finite temperature. There is a gap of $2m$ in the real part of conductivity as can be seen from the Heaviside function in Eq. \ref{eq:massive_scalar_conductivity_real} (also see Fig. \ref{fig:2d_real}). This gap is the minimum energy required to create a particle-antiparticle pair in the system. In the special case of d = 2 and T = 0, the optical conductivity can be calculated analytically,
\beq
\sigma_1(\omega) &=& \frac{g^2}{16} \theta(\omega^2 - (2m)^2)(1-\frac{(2m)^2}{\omega^2}) \label{eq:scalar_conductivity_d2T0_real} \\
\sigma_2(\omega) &=& \frac{g^2}{16\pi}(-\frac{4m}{\omega} + (1-\frac{(2m)^2}{\omega^2})\ln\big|\frac{\omega-2m}{\omega+2m}\big| ). \label{eq:scalar_conductivity_d2T0_imag}
\eeq
The plots of the real and imaginary parts of conductivity in this case are shown in Fig. \ref{fig:2d}.

\section{Unparticle in Matsubara Space} \label{app:unparticle_matsubara}
The U(1) invariant unparticle effective action in d+1 Matsubara space is given by
\begin{align}  \label{eq:unparticle_action}
S = \ &T\sum\limits_{m}\int \frac{d^dp}{(2\pi)^d}\phi^\dagger_m(p)G^{-1}_{m}(p)\phi_m(p) \nonumber \\
& + T^2\sum\limits_{m,n}\int \frac{d^dpd^dq}{(2\pi)^{2d}}\phi^\dagger_{m+n}(p+q)\phi_m(p)A_{\mu,n}(q)g\Gamma^\mu_{m,n}(p,q) \nonumber \\
& + T^3\sum\limits_{m,n_1,n_2}\int \frac{d^dpd^dq_1d^dq_2}{(2\pi)^{3d}}\phi^\dagger_{m+n_1+n_2}(p+q_1+q_2)\phi_m(p)A_{\mu,n_1}(q_1)A_{\nu,n_2}(q_2)\frac{g^2}{2}\Gamma^{\mu\nu}_{m,n_1,n_2}(p,q_1,q_2),
\end{align} 
where subscripts of the fields, the propagators, and the vertices denote the dependence on the bosonic Matsubara frequency ($\omega_n = \frac{2\pi n}{\beta}$ with $n$ being an integer). The propagator is
\beq \label{eq:unparticle_propagator}
G_n(k) &=& -\frac{A_{d_U}}{2\sin(d_U\pi)}\frac{1}{(k^2+\omega_n^2)^{\frac{d+1}{2}-d_U}} .
\eeq
The two unparticles one gauge field vertex is
\begin{align} \label{eq:unparticle_vertex1}
\Gamma^\mu_{m,n}(p,q) = (2p^\mu+q^\mu)\mathcal{F}_{m,n}(p,q)
\end{align}
and the two unparticles two gauge field vertex is
\begin{align} \label{eq:unparticle_vertex2}
\Gamma^{\mu\nu}_{m,n_1,n_2}(p,q_1,q_2) = &2\delta^{\mu\nu}\mathcal{F}_{m,n_1+n_2}(p,q_1+q_2)  \nonumber \\
&+\frac{(2p+q_2)^\nu(2p+2q_2+q_1)^\mu}{q_1^2+2(p+q_2)\cdot q_1+\omega_{n_1}^2+2(\omega_m+\omega_{n_2})\omega_{n_1}}(\mathcal{F}_{m,n_1+n_2}(p,q_1+q_2)-\mathcal{F}_{m,n_2}(p,q_2)) \nonumber \\
&+\frac{(2p+q_1)^\mu(2p+2q_1+q_2)^\nu}{q_2^2+2(p+q_1)\cdot q_2+\omega_{n_2}^2+2(\omega_m+\omega_{n_1})\omega_{n_2}}(\mathcal{F}_{m,n_1+n_2}(p,q_1+q_2)-\mathcal{F}_{m,n_1}(p,q_1)),
\end{align}
where the function $\mathcal{F}(p,q)$ is
\beq
\mathcal{F}_{m,n}(p,q)  = \frac{G^{-1}_{m+n}(p+q) - G^{-1}_m(p)}{q^2+2p\cdot q+\omega_n^2+2\omega_m\omega_n}.
\eeq
From the expression $p^\mu$, $p^0$ is Matsubara frequency and $p^i$, for $i$ = 1 to d, is an ith component of d-dimensional momentum. 

\section{Calculation of Optical Conductivity from Unparticle Effective Action} \label{app:summation_matsubara}
From Eq. \ref{eq:response_general}, we refer to the first term as the diamanetic term, $K^{\mu\nu}_{dia,n}(q)$, and the second term as the paramagnetic term, $K^{\mu\nu}_{para,n}(q)$.
Substituting Eqs. \ref{eq:unparticle_propagator}, \ref{eq:unparticle_vertex1}, and \ref{eq:unparticle_vertex2} into these two terms, one obtains
\begin{align} \label{eq:kdia}
K^{\mu\nu}_{dia,n}(q) & = 2g^2T\sum\limits_{m}\int \frac{d^dp}{(2\pi)^d}\Bigg( (\frac{d+1}{2}-d_U)\frac{\delta^{\mu\nu}}{(\omega_m^2+p^2)} \nonumber \\
& \tabii
- \frac{(2p+q)^\nu(2p+q)^\mu}{q^2+2p\cdot q+\omega_{n}^2+2\omega_m\omega_{n}}\frac{(\frac{d+1}{2}-d_U)}{\omega_m^2+p^2} \nonumber \\
& \tabii
-\frac{(2p+q)^\nu(2p+q)^\mu}{(q^2+2p\cdot q+\omega_n^2+2\omega_m\omega_n)^2} \nonumber \\
& \tabii
+ \frac{(2p+q)^\nu(2p+q)^\mu}{(q^2+2p\cdot q+\omega_n^2+2\omega_m\omega_n)^2}\bigg(\frac{(p+q)^2+(\omega_m+\omega_n)^2}{p^2+\omega_m^2}\bigg)^{\frac{d+1}{2}-d_U}\Bigg)
\end{align}
and
\begin{align} \label{eq:kpara}
K^{\mu\nu}_{para,n}(q)= 2g^2T\sum\limits_{m}\int \frac{d^dp}{(2\pi)^d} \Bigg(& \frac{(2p^\mu+q^\mu)(2p^\nu+q^\nu)}{(q^2+2p\cdot q+\omega_m^2+2\omega_m\omega_n)^2} \nonumber \\
&
-\frac{1}{2} \frac{(2p^\mu+q^\mu)(2p^\nu+q^\nu)}{(q^2+2p\cdot q+\omega_m^2+2\omega_m\omega_n)^2}\Bigg(\frac{(p+q)^2+(\omega_m+\omega_n)^2}{p^2+\omega_m^2}\Bigg)^{\frac{d+1}{2}-d_U} \nonumber \\
& -\frac{1}{2} \frac{(2p^\mu+q^\mu)(2p^\nu+q^\nu)}{(q^2+2p\cdot q+\omega_m^2+2\omega_m\omega_n)^2}\Bigg(\frac{p^2+\omega_m^2}{(p+q)^2+(\omega_m+\omega_n)^2}\Bigg)^{\frac{d+1}{2}-d_U} \Bigg).
\end{align}
In obtaining the result for the diamagnetic term, we interpret $\mathcal{F}_{m,0}(p,0)$ as
\beq
\lim\limits_{q\rightarrow0,n\rightarrow0}\mathcal{F}_{m,n}(p,q) &=& \frac{d}{dx^2}(x^2)^{\frac{d+1}{2}-d_U}\bigg\vert_{x^2=p^2+\omega_m^2} \nonumber \\
&=& (\frac{d+1}{2}-d_U)(p^2+\omega_m^2)^{\frac{d-1}{2}-d_U}.
\eeq
We split $K_{dia}$ into four terms and $K_{para}$ into three terms:
\beq
K^{\mu\nu}_{dia1,n}(q) &=& 2g^2\bigg( \frac{d+1}{2}-d_U \bigg)T\sum\limits_{m}\int \frac{d^dp}{(2\pi)^d} \frac{\delta^{\mu\nu}}{\omega_m^2+p^2}  \label{eq:Kd1} \\
K^{\mu\nu}_{dia2,n}(q) &=& -2g^2\bigg( \frac{d+1}{2}-d_U \bigg)T\sum\limits_{m}\int \frac{d^dp}{(2\pi)^d}\frac{(2p+q)^\nu(2p+q)^\mu}{q^2+2p\cdot q+\omega_{n}^2+2\omega_m\omega_{n}}\frac{1}{\omega_m^2+p^2} \label{eq:Kd2} \\
K^{\mu\nu}_{dia3,n}(q) &=& -2g^2T\sum\limits_{m}\int \frac{d^dp}{(2\pi)^d}\frac{(2p+q)^\nu(2p+q)^\mu}{(q^2+2p\cdot q+\omega_n^2+2\omega_m\omega_n)^2} \label{eq:Kd3} \\
K^{\mu\nu}_{dia4,n}(q) &=& 2g^2T\sum\limits_{m}\int \frac{d^dp}{(2\pi)^d} \frac{(2p+q)^\nu(2p+q)^\mu}{(q^2+2p\cdot q+\omega_n^2+2\omega_m\omega_n)^2}\bigg(\frac{(p+q)^2+(\omega_m+\omega_n)^2}{p^2+\omega_m^2}\bigg)^{\frac{d+1}{2}-d_U} \label{eq:Kd4} \\
K^{\mu\nu}_{para1,n}(q) &=& -K^{\mu\nu}_{dia3,n}(q) \label{eq:Kp1}  \\
K^{\mu\nu}_{para2,n}(q) &=& -\frac{1}{2}K^{\mu\nu}_{dia4,n}(q) \label{eq:Kp2}  \\
K^{\mu\nu}_{para3,n}(q) &=& -g^2T\sum\limits_{m}\int \frac{d^dp}{(2\pi)^d} \frac{(2p+q)^\nu(2p+q)^\mu}{(q^2+2p\cdot q+\omega_n^2+2\omega_m\omega_n)^2}\Bigg(\frac{p^2+\omega_m^2}{(p+q)^2+(\omega_m+\omega_n)^2}\Bigg)^{\frac{d+1}{2}-d_U}. \label{eq:Kp3}
\eeq
We show here the Matsubara summation using the contour integral calculation technique on these seven terms.

\subsection{$K_{dia1}$}
\begin{figure}
        \centering
        \subfigure[$\ K_{dia1}$ \label{fig:Kd1}]{\includegraphics[scale=0.5]{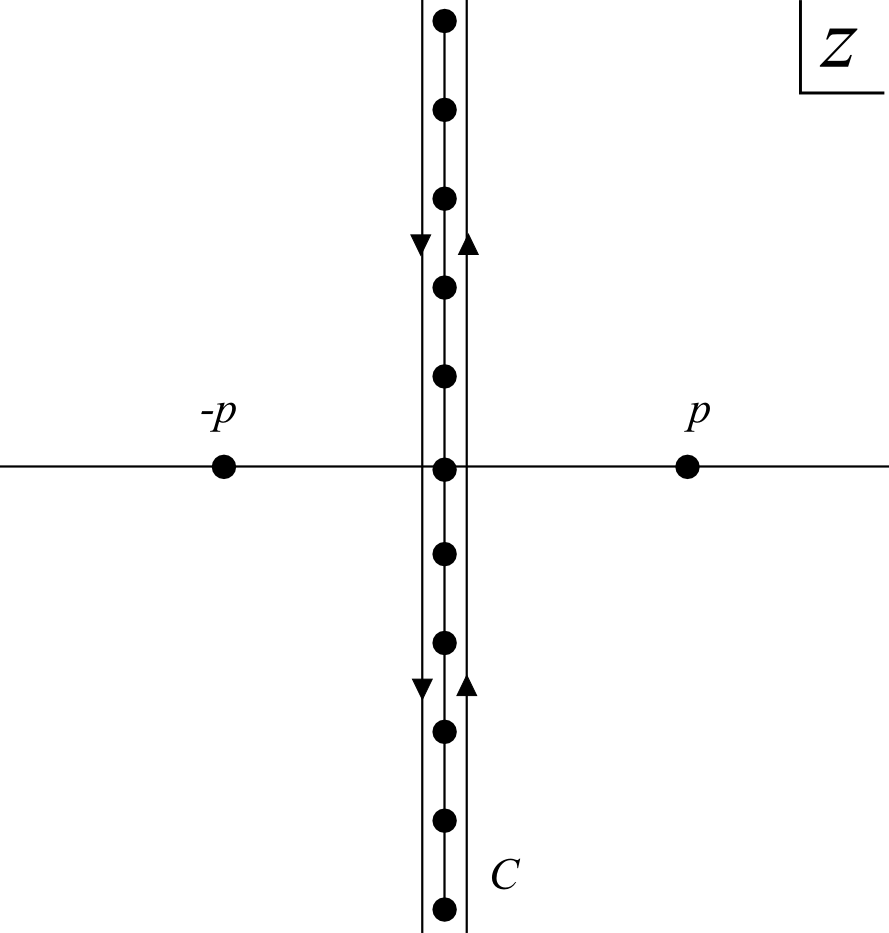}}
        \subfigure[$\ K_{dia2}$ \label{fig:Kd2_odd}]{\includegraphics[scale=0.5]{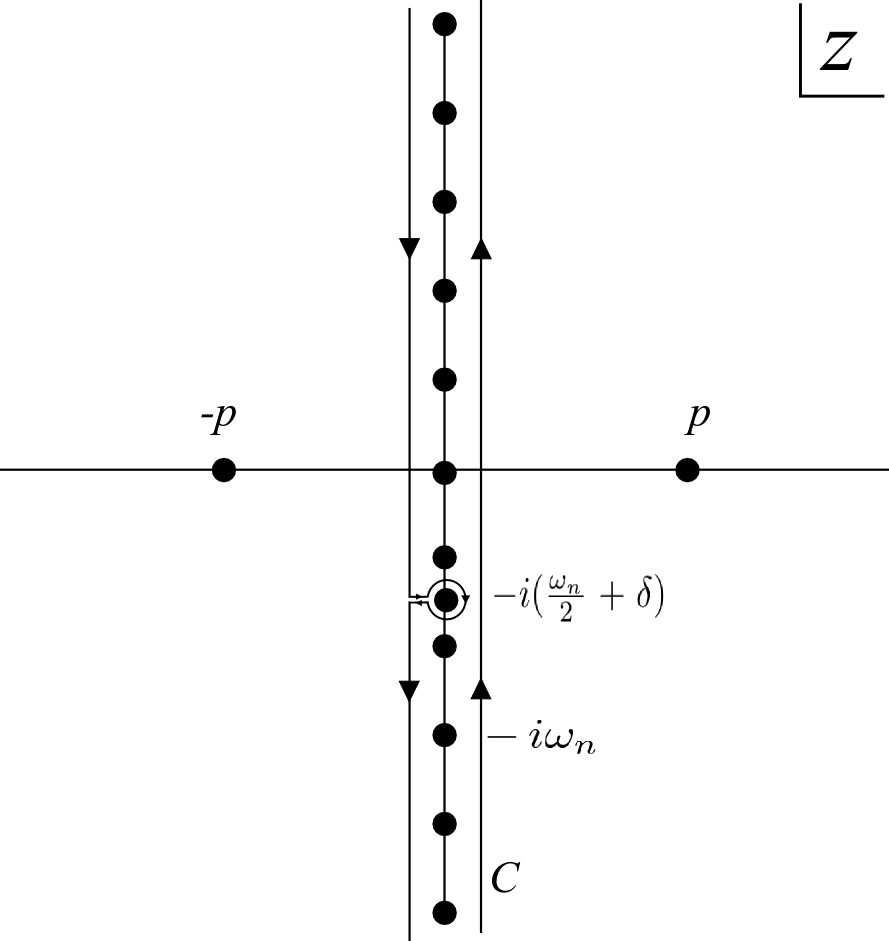}}
        \subfigure[$\ K_{dia3}$ \label{fig:Kd3_odd}]{\includegraphics[scale=0.5]{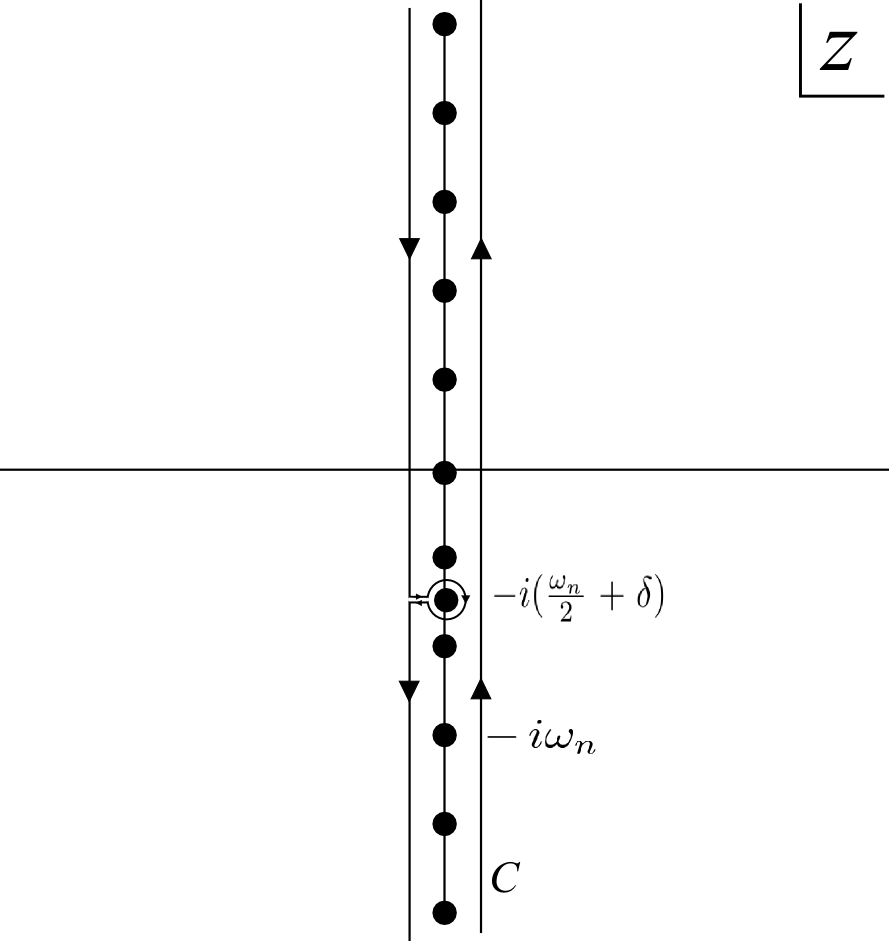}}
		\caption{The contours used in the summation of three diamagnetic terms. }
\end{figure}

The $ii$ component of $K_{dia1}$ in the $q\rightarrow0$ limit is
\beq
K^{ii}_{dia1,n}(q\rightarrow0) = 2g^2\bigg( \frac{d+1}{2}-d_U \bigg)T\sum\limits_{m}\int \frac{d^dp}{(2\pi)^d} \frac{1}{\omega_m^2+p^2}. \nonumber
\eeq
Inserting a factor $1 = \frac{dp_i}{dp_i}$ into the integrand and then performing an integration by parts, we have
\beq
K^{ii}_{dia1,n}(q\rightarrow0) = 2g^2\bigg( \frac{d+1}{2}-d_U \bigg)T\sum\limits_{m}\int \frac{d^dp}{(2\pi)^d} \frac{2p^2_i}{(\omega_m^2+p^2)^2}. \nonumber
\eeq
Carrying out the Matsubara summation using the the contour shown in Fig. \ref{fig:Kd1}, one obtains
\beq
T\sum\limits_{m}\frac{1}{(\omega_m^2+p^2)^2} &=& T\sum\limits_{m}\frac{1}{(\omega_m^2+p^2)^2} \nonumber \\
&=& \frac{\beta}{2p^2}N(p)(N(p)+1) + \frac{1}{2p^3}g(p), \nonumber
\eeq

where $g(x) = \frac{1}{2}\coth(\frac{\beta x}{2})$ and $N(x) = \frac{1}{e^{\beta x}-1}$ is the Bose-Einstein distribution. The result is
\beq
K^{ii}_{dia1,n}(q\rightarrow0) = 2g^2\bigg( \frac{d+1}{2}-d_U \bigg)\int \frac{d^dp}{(2\pi)^d}p^2_i \bigg(\frac{\beta}{p^2}N(p)(N(p)+1) + \frac{1}{p^3}g(p) \bigg).
\eeq

\subsection{$K_{dia2}$}
We rewrite the $ii$ components of Eq. \ref{eq:Kd2} as
\begin{align}
&K^{ii}_{dia2,n}(q) = \frac{-ig^2}{\omega_n}\bigg( \frac{d+1}{2}-d_U \bigg)T\sum\limits_{m}\int \frac{d^dp}{(2\pi)^d}(2p^i+q^i)^2  \frac{1}{i\omega_m+i(\delta+\frac{\omega_{n}}{2})}\frac{1}{\omega_m^2+p^2}, \nonumber
\end{align}
where $\delta$ is defined as
\beq
2\omega_n\delta = q^2+2p\cdot q. \label{eq:delta}
\eeq
We will not take the limit $\delta\rightarrow0$ (or $q\rightarrow0$) right away, but we will do so after combining the contributions from Eq. \ref{eq:Kd2} - \ref{eq:Kd4}. The reason for this is given below. We convert the summation into the contour integral with its contour shown in Fig. \ref{fig:Kd2_odd}:
\beq
T\sum\limits_{m}\frac{1}{i\omega_m+i(\delta+\frac{\omega_{n}}{2})}\frac{1}{\omega_m^2+p^2} = 
-\int_C dz g(z)\frac{1}{z+i(\frac{\omega_{n}}{2}+\delta)}\frac{1}{z^2-p^2}. \nonumber
\eeq

In addition to the poles of $g(z)$, there are poles at $\pm p$ and $-i(\frac{\omega_{n}}{2}+\delta)$. Had we set $\delta = 0$, there would be a pole at $-i\frac{\omega_{n}}{2}$. This is problematic. The reason is that when $n$ is odd, the integrand has a simple pole at $-i\frac{\omega_{n}}{2}$. But when $n$ is even, the integrand has a double pole at $-i\frac{\omega_{n}}{2}$, because $\frac{\omega_{n}}{2}$ would correspond to another Matsubara frequency $\omega_l$ with $l = \frac{n}{2}$. This problem is related to the fact that the summation $T\sum\limits_{m}\frac{1}{i\omega_m+i\frac{\omega_{n}}{2}}\frac{1}{\omega_m^2+p^2}$ diverges when $n$ is even. Hence, we need to keep $\delta$ finite and then take the limit $\delta\rightarrow 0$ after we combine Eqs. \ref{eq:Kd2} - \ref{eq:Kd4}. Deforming the contour to infinty and then summing the residues at $\pm p$ and $-i(\frac{\omega_{n}}{2}+\delta)$, one finds
\begin{align}
K^{ii}_{dia2,n}(q) =  &\frac{-ig^2}{\omega_n}\bigg( \frac{d+1}{2}-d_U \bigg)\int \frac{d^dp}{(2\pi)^d}(2p^i+q^i)^2  \bigg(-\frac{i}{2}\frac{\cot(\frac{\beta}{2}(\delta+\frac{\omega}{2}))}{p^2+(\delta+\frac{\omega_n}{2})^2}+\frac{g(p)}{2p(p+i(\delta+\frac{\omega_n}{2}))}+\frac{g(-p)}{2p(p-i(\delta+\frac{\omega_n}{2}))} \bigg). \label{eq:kd2_f}
\end{align}

\subsection{$K_{dia3}$}
We rewrite Eq. \ref{eq:Kd3} using the definition of $\delta$ given in Eq. \ref{eq:delta} and then perform the Matsubara summation with the contour given in Fig. \ref{fig:Kd3_odd}:
\beq
K^{ii}_{dia3,n}(q) &=&  \frac{g^2}{2\omega_n^2}T\sum\limits_{m}\int \frac{d^dp}{(2\pi)^d}(2p^i+q^i)^2  \frac{1}{(i\omega_m+i(\delta+\frac{\omega_n}{2}))^2} \nonumber \\
&=&  \frac{-\beta g^2}{8\omega_n^2}\int \frac{d^dp}{(2\pi)^d}(2p^i+q^i)^2\csc^2\bigg(\frac{\beta}{2}(\delta+\frac{\omega_n}{2}) \bigg). \label{eq:kd3_f}
\eeq

\subsection{$K_{dia4}$}
\begin{figure}[h!] 
\centering
\includegraphics[width=0.4\textwidth]{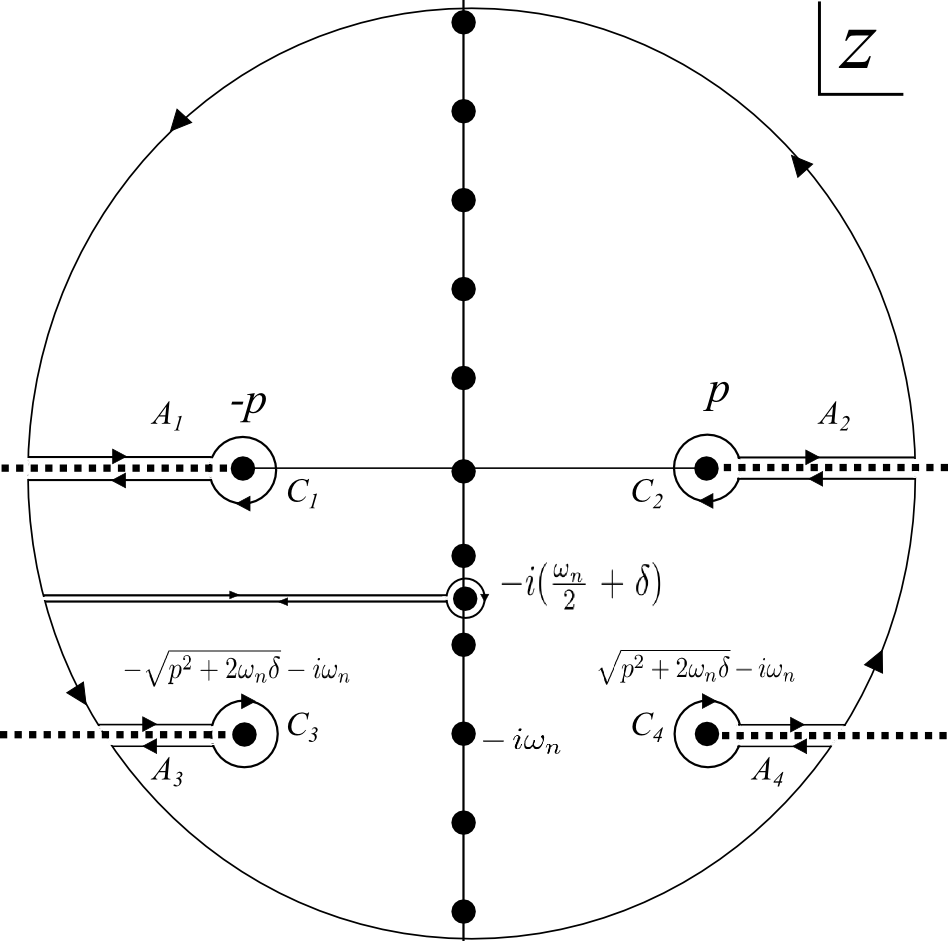}
\caption{The contour of $K_{dia4}$ in the case of odd $n$ and small $\delta$. The paths that go along the branch cuts are $A_1$ to $A_4$ and the paths that go around the branch points are $C_1$ to $C_4$.}  \label{fig:Kd4_odd}
\end{figure}

We rewrite Eq. \ref{eq:Kd4} with the definition of $\delta$ given in Eq. \ref{eq:delta} and then convert it into the contour integral with the contour shown in Fig. \ref{fig:Kd4_odd}:
\beq
K^{\mu\nu}_{dia4,n}(q) &=& -\frac{g^2}{2\omega_n^2}\int \frac{d^dp}{(2\pi)^d} (2p^i+q^i)^2T\sum\limits_{m}\frac{1}{(i\omega_m+i(\delta+\frac{\omega_n}{2}))^2}\bigg(\frac{p^2 + 2\omega_n\delta -(i\omega_m+i\omega_n)^2}{p^2+\omega_m^2}\bigg)^{\frac{d+1}{2}-d_U} \nonumber  \\
&=& -\frac{g^2}{2\omega_n^2}\int \frac{d^dp}{(2\pi)^d} (2p^i+q^i)^2\frac{1}{2\pi i}\int_C dz \frac{1}{(z+i(\delta+\frac{\omega_n}{2}))^2}\bigg(\frac{p^2 + 2\omega_n\delta-(z+i\omega_n)^2}{p^2-z^2}\bigg)^{\frac{d+1}{2}-d_U}. \label{eq:kd_4_contour} 
\eeq
In addition to the pole at $-i(\delta+\frac{\omega}{2})$, there are branch points at $\pm p$ and $\pm \sqrt{p^2+2\omega_n\delta} - i\omega_n$. Thus, we need to consistently choose the Riemann surface that will be used in the contour integration. We split up the term with power $\frac{d+1}{2}-d_U$ in the integrand as follows,
\begin{align}
\bigg(\frac{p^2 + 2\omega_n\delta-(z+i\omega_n)^2}{p^2-z^2}\bigg)^{\frac{d+1}{2}-d_U} &= \bigg(\frac{(z+i\omega_n)^2-(p^2 + 2\omega_n\delta)}{z^2-p^2}\bigg)^{\frac{d+1}{2}-d_U} \nonumber \\
&= \frac{(z+i\omega_n-\sqrt{p^2 + 2\omega_n\delta})^{\frac{d+1}{2}-d_U}(z+i\omega_n+\sqrt{p^2 + 2\omega_n\delta})^{\frac{d+1}{2}-d_U}}{(z-p)^{\frac{d+1}{2}-d_U}(z+p)^{\frac{d+1}{2}-d_U}}. \nonumber
\end{align}
The constraint on the choice of the Riemann surface is the condition that the contour integral is real. In other words, we require the residue at $z^* = i\omega_m$ of Eq. \ref{eq:kd_4_contour}, that is,
$
\frac{(i\omega_m+i\omega_n-\sqrt{p^2 + 2\omega_n\delta})^{\frac{d+1}{2}-d_U}(i\omega_m+i\omega_n+\sqrt{p^2 + 2\omega_n\delta})^{\frac{d+1}{2}-d_U}}{(i\omega_m-p)^{\frac{d+1}{2}-d_U}(i\omega_m+p)^{\frac{d+1}{2}-d_U}} \nonumber
$, to be real. \\
 \\
We choose branch cuts as in Fig. \ref{fig:Kd4_odd}. One choice of the definitions of the phase angles that are satisfied by the reality constraint mentioned above is listed below. \\ \\
For the term $(z+p)^{\frac{d+1}{2}-d_U}$, the definition of the phase angle is  $-\pi \leq \theta_1 <  \pi$. \\ \\
For the term $(z-p)^{\frac{d+1}{2}-d_U}$, the definition of the phase angle is $0 \leq \theta_2 < 2\pi$. \\ \\
For the term $(z +i\omega_n+\sqrt{p^2+2\delta\omega_n})^{\frac{d+1}{2}-d_U}$, the definition of the phase angle is  $-\pi \leq \theta_3 <  \pi$. \\ \\
Lastly, the phase angle is $0 \leq \theta_4 <  2\pi$ in the term $(z -i\omega_n-\sqrt{p^2+2\delta\omega_n})^{\frac{d+1}{2}-d_U}$. \\ \\
The contour integral can be split into the contribution from the residue at  $-i(\delta+\frac{\omega}{2})$, the contribution from the discontinuity across the branch cuts ($A_i$'s in Fig. \ref{fig:Kd4_odd}), and the contribution from the small circles around the branch points ($C_i$'s in Fig. \ref{fig:Kd4_odd}). The residue contribution to the Matsubara summation is
\begin{align}
&-Res[\frac{1}{(z+i(\delta+\frac{\omega_n}{2}))^2}\bigg(\frac{p^2 + 2\omega_n\delta-(z+i\omega_n)^2}{p^2-z^2}\bigg)^{\frac{d+1}{2}-d_U}] = -\frac{\beta}{4}\csc^2\bigg(\frac{\beta}{2}(\delta+\frac{\omega_n}{2})\bigg) - \omega_n(\frac{d+1}{2}-d_U)\frac{\cot(\frac{\beta}{2}(\delta+\frac{\omega_n}{2}))}{p^2+(\delta+\frac{\omega_n}{2})^2}.
\end{align}
Hence, the residue contribution to $K_{dia4}$ is
\beq
&& K^{ii}_{dia4,res,n}(q) = \frac{g^2}{2\omega_n^2}\int \frac{d^dp}{(2\pi)^d} (2p^i+q^i)^2 \bigg( \frac{\beta}{4}\csc^2\bigg(\frac{\beta}{2}(\delta+\frac{\omega_n}{2})\bigg) + \omega_n(\frac{d+1}{2}-d_U)\frac{\cot(\frac{\beta}{2}(\delta+\frac{\omega_n}{2}))}{p^2+(\delta+\frac{\omega_n}{2})^2}\bigg).\label{eq:kd4_res_f}
\eeq
For the contributions from the discontinuity across the brach cuts and the small circles around the branch points, we take the limit $q\rightarrow0$ or $\delta\rightarrow0$ to simplify the calculation. Let us consider the contribution from the discontinuities across the branch cuts (the integral along $A_1$ to $A_4$):
\begin{align}
=& \frac{1}{2\pi i}\int\limits_{-\infty}^{-p-\varepsilon}dz \frac{g(z)}{(z+i\frac{\omega_n}{2})^2}\frac{(z+i\omega_n-p)^{\frac{d+1}{2}-d_U}(z+i\omega_n+p)^{\frac{d+1}{2}-d_U}}{(z-p)^{\frac{d+1}{2}-d_U}} \bigg(\frac{1}{(z_{+}+p)^{\frac{d+1}{2}-d_U}}-\frac{1}{(z_{-}+p)^{\frac{d+1}{2}-d_U}}\bigg) \nonumber \\
&+\frac{1}{2\pi i}\int\limits_{p+\varepsilon}^{\infty}dz \frac{g(z)}{(z+i\frac{\omega_n}{2})^2}\frac{(z+i\omega_n-p)^{\frac{d+1}{2}-d_U}(z+i\omega_n+p)^{\frac{d+1}{2}-d_U}}{(z+p)^{\frac{d+1}{2}-d_U}} \bigg(\frac{1}{(z_{+}-p)^{\frac{d+1}{2}-d_U}}-\frac{1}{(z_{-}-p)^{\frac{d+1}{2}-d_U}}\bigg) \nonumber \\
&+\frac{1}{2\pi i}\int\limits_{-i\omega_n-\infty}^{-i\omega_n-p-\varepsilon}dz \frac{g(z)}{(z+i\frac{\omega_n}{2})^2}\frac{(z+i\omega_n-p)^{\frac{d+1}{2}-d_U}}{(z-p)^{\frac{d+1}{2}-d_U}(z+p)^{\frac{d+1}{2}-d_U}} \bigg((z_{+}+i\omega_n+p)^{\frac{d+1}{2}-d_U}-(z_{-}+i\omega_n+p)^{\frac{d+1}{2}-d_U}\bigg) \nonumber \\
&+\frac{1}{2\pi i}\int\limits_{-i\omega_n+p+\varepsilon}^{-i\omega_n+\infty}dz \frac{g(z)}{(z+i\frac{\omega_n}{2})^2}\frac{(z+i\omega_n+p)^{\frac{d+1}{2}-d_U}}{(z-p)^{\frac{d+1}{2}-d_U}(z+p)^{\frac{d+1}{2}-d_U}} \bigg((z_{+}+i\omega_n-p)^{\frac{d+1}{2}-d_U}-(z_{-}+i\omega_n-p)^{\frac{d+1}{2}-d_U}\bigg), \nonumber
\end{align}
where $z_{\pm} = z\pm i\eta$ with $\eta\rightarrow0^+$. Using the definitions of the angles $\theta_i$ defined above, we find that the discontinuities across the four cuts are
\begin{align}
\frac{1}{(z_{+}+p)^{\frac{d+1}{2}-d_U}}-\frac{1}{(z_{-}+p)^{\frac{d+1}{2}-d_U}} & = \frac{-2i\sin{\pi(\frac{d+1}{2}-d_U)}}{|z+p|^{\frac{d+1}{2}-d_U}} \label{eq:cut1} \\
\frac{1}{(z_{+}-p)^{\frac{d+1}{2}-d_U}}-\frac{1}{(z_{-}-p)^{\frac{d+1}{2}-d_U}} & = \frac{2i\sin{\pi(\frac{d+1}{2}-d_U)}e^{-i\pi(\frac{d+1}{2}-d_U)}}{|z-p|^{\frac{d+1}{2}-d_U}} \label{eq:cut2} \\
(z_{+}+i\omega_n+p)^{\frac{d+1}{2}-d_U}-(z_{-}+i\omega_n+p)^{\frac{d+1}{2}-d_U} & = 2i\sin{\pi(\frac{d+1}{2}-d_U)}|z+i\omega_n+p|^{\frac{d+1}{2}-d_U} \label{eq:cut3} \\
(z_{+}+i\omega_n-p)^{\frac{d+1}{2}-d_U}-(z_{-}+i\omega_n-p)^{\frac{d+1}{2}-d_U} & = -2i\sin{\pi(\frac{d+1}{2}-d_U)}e^{i\pi(\frac{d+1}{2}-d_U)}|z+i\omega_n-p|^{\frac{d+1}{2}-d_U}. \label{eq:cut4}
\end{align}
Substiuting the four discontinuities into the integral along $A_1$ to $A_4$ and then changing variables so that the limit of integration is from $\varepsilon$ to $\infty$ yields
\begin{align}
\frac{\sin{\pi(\frac{d+1}{2}-d_U)}}{\pi} \int\limits_{\varepsilon}^{\infty}dz \Bigg(&
\frac{g(z+p)}{(-z+i\frac{\omega_n}{2}-p)^2}\frac{(-z+i\omega_n-2p)^{\frac{d+1}{2}-d_U}(-z+i\omega_n)^{\frac{d+1}{2}-d_U}}{(-z-2p)^{\frac{d+1}{2}-d_U}z^{\frac{d+1}{2}-d_U}}  \nonumber \\
&  +\frac{g(z+p)}{(z+i\frac{\omega_n}{2}+p)^2}\frac{(z+i\omega_n)^{\frac{d+1}{2}-d_U}(z+i\omega_n+2p)^{\frac{d+1}{2}-d_U}}{z^{\frac{d+1}{2}-d_U}(z+2p)^{\frac{d+1}{2}-d_U}}e^{-i\pi(\frac{d+1}{2}-d_U)}   \nonumber \\
&  -\frac{g(z+p)}{(-z-i\frac{\omega_n}{2}-p)^2}\frac{(-z-2p)^{\frac{d+1}{2}-d_U}z^{\frac{d+1}{2}-d_U}}{(-z-i\omega_n-2p)^{\frac{d+1}{2}-d_U}(-z-i\omega_n)^{\frac{d+1}{2}-d_U}}  \nonumber \\
&  -\frac{g(z+p)}{(z-i\frac{\omega_n}{2}+p)^2}\frac{z^{\frac{d+1}{2}-d_U}(z+2p)^{\frac{d+1}{2}-d_U}}{(z-i\omega_n)^{\frac{d+1}{2}-d_U}(z-i\omega_n+2p)^{\frac{d+1}{2}-d_U}} e^{i\pi(\frac{d+1}{2}-d_U)}.
\Bigg). \nonumber
\end{align}
On the third and fourth lines, we use the fact that $g(z+i\omega_n) = \frac{1}{2}\coth(\frac{\beta (z+i\omega_n)}{2}) = g(z)$ and, on the first and the third lines, we use $g(-z) = -g(z)$ . To further simplify the integral, we pull factors of $-1$ out from the terms with power of $\frac{d+1}{2}-d_U$ and -1 in front of z to make the coefficients in front of z all be +1. At the same time, we redefine all the angles $\theta_i$ to be evaluated by the power $\frac{d+1}{2}-d_U$ to be in one range, i.e. $-\pi <\theta_i < \pi$. These operations amount to multiplying each line by a factor $e^{im_j\pi(\frac{d+1}{2}-d_U)}$ (subscript j means line j). We find that $m_1 = 1$, $m_2 = 0$, $m_3 = 1$, and $m_4 = -2$. The result is
\begin{align}
\frac{\sin{\pi(\frac{d+1}{2}-d_U)}}{\pi} \int\limits_{\varepsilon}^{\infty}dz \Bigg(&
\frac{g(z+p)}{(z-i\frac{\omega_n}{2}+p)^2}\frac{(z-i\omega_n+2p)^{\frac{d+1}{2}-d_U}(z-i\omega_n)^{\frac{d+1}{2}-d_U}}{(z+2p)^{\frac{d+1}{2}-d_U}z^{\frac{d+1}{2}-d_U}} e^{i\pi(\frac{d+1}{2}-d_U)} \nonumber \\
&  +\frac{g(z+p)}{(z+i\frac{\omega_n}{2}+p)^2}\frac{(z+i\omega_n)^{\frac{d+1}{2}-d_U}(z+i\omega_n+2p)^{\frac{d+1}{2}-d_U}}{z^{\frac{d+1}{2}-d_U}(z+2p)^{\frac{d+1}{2}-d_U}} e^{-i\pi(\frac{d+1}{2}-d_U)}  \nonumber \\
&  -\frac{g(z+p)}{(z+i\frac{\omega_n}{2}+p)^2}\frac{(z+2p)^{\frac{d+1}{2}-d_U}z^{\frac{d+1}{2}-d_U}}{(z+i\omega_n+2p)^{\frac{d+1}{2}-d_U}(z+i\omega_n)^{\frac{d+1}{2}-d_U}} e^{i\pi(\frac{d+1}{2}-d_U)} \nonumber \\
&  -\frac{g(z+p)}{(z-i\frac{\omega_n}{2}+p)^2}\frac{z^{\frac{d+1}{2}-d_U}(z+2p)^{\frac{d+1}{2}-d_U}}{(z-i\omega_n)^{\frac{d+1}{2}-d_U}(z-i\omega_n+2p)^{\frac{d+1}{2}-d_U}} e^{-i\pi(\frac{d+1}{2}-d_U)}
\Bigg). \nonumber
\end{align}
Note that the first and the third terms are complex conjugates of the second and the fourth terms, respectively, and hence the result of the Matsubara summation is real as we suspected.
Finally, let us consider the contribution from the integrals along the small circles around the branch points ($C_1$ to $C_4$):
\beq
\frac{1}{2\pi i}\int_{C_i}\frac{g(z)}{(z+i\frac{\omega_n}{2})^2}\bigg(\frac{(z+i\omega_n)^2-p^2}{z^2-p^2}\bigg)^{\frac{d+1}{2}-d_U}.
\eeq
For the curves $C_1$ and $C_2$, we let $z = \mp p + \varepsilon e^{i\theta}$, where $-$ refers to $C_1$ and $+$ refers to $C_2$. The integral is along the clockwise direction; hence it needs to be multiplied by a factor of $-1$. The result is
\beq 
-\frac{\varepsilon^{\frac{1-d}{2}+d_U}}{2\pi} \int\limits_{\theta_i}^{\theta_f}d\theta e^{i\theta} \frac{g(\mp p + \varepsilon e^{i\theta})}{(\mp p + i\frac{\omega_n}{2}+\varepsilon e^{i\theta})^2} \bigg(\frac{(\mp 2p+i\omega_n+\varepsilon e^{i\theta})(i\omega_n+\varepsilon e^{i\theta})}{(\mp 2p + \varepsilon e^{i\theta})e^{i\theta}}\bigg)^{\frac{d+1}{2}-d_U}. \label{eq:c1c2}
\eeq
In the case of curve $C_3$ and $C_4$, we let $z = \mp p - i\omega_n + \varepsilon e^{i\theta}$, where $-$ is for $C_3$ and $+$ is for $C_4$. The result is
\beq
= -\frac{\varepsilon^{\frac{d+3}{2}-d_U}}{2\pi} \int\limits_{\theta_i}^{\theta_f}d\theta e^{i\theta} \frac{g(\mp p + \varepsilon e^{i\theta})}{(\mp p - i\frac{\omega_n}{2}+\varepsilon e^{i\theta})^2}
\bigg(\frac{(\mp 2p+\varepsilon e^{i\theta})e^{i\theta}}{(\mp 2p -i\omega_n+ \varepsilon e^{i\theta})(-i\omega_n + \varepsilon e^{i\theta})}\bigg)^{\frac{d+1}{2}-d_U}. \label{eq:c3c4}
\eeq
Eqs. \ref{eq:c1c2} and \ref{eq:c3c4} vanish in the limit $\varepsilon\rightarrow0$ when the exponents of $\varepsilon^{\frac{d+3}{2}-d_U}$ and $\varepsilon^{\frac{1-d}{2}+d_U}$ are greater than zero, that is, when $\frac{d-1}{2}<d_U<\frac{d+3}{2}$. From the unitarity bound, $d_U > \frac{d-1}{2}$. If we now also require that $d_U<\frac{d+3}{2}$, we find that there will be no contributions from the $C_i$'s.

Hence, the contribution from the branch cuts (both $A_i$'s and $C_i$'s) to $K_{dia4}$ in the limit $q\rightarrow0$ is
\begin{align}
K^{ii}_{dia4,cut,n}(q\rightarrow0) & = -g^2\frac{\sin{\pi(\frac{d+1}{2}-d_U)}}{2\pi\omega_n^2} \int \frac{d^dp}{(2\pi)^d}(2p^i+q^i)^2 \int\limits_{\varepsilon}^{\infty}dz  \nonumber \\
\Bigg(
&
\frac{g(z+p)}{(z-i\frac{\omega_n}{2}+p)^2}\frac{(z-i\omega_n+2p)^{\frac{d+1}{2}-d_U}(z-i\omega_n)^{\frac{d+1}{2}-d_U}}{(z+2p)^{\frac{d+1}{2}-d_U}z^{\frac{d+1}{2}-d_U}} e^{i\pi(\frac{d+1}{2}-d_U)} \nonumber \\
&  +\frac{g(z+p)}{(z+i\frac{\omega_n}{2}+p)^2}\frac{(z+i\omega_n)^{\frac{d+1}{2}-d_U}(z+i\omega_n+2p)^{\frac{d+1}{2}-d_U}}{z^{\frac{d+1}{2}-d_U}(z+2p)^{\frac{d+1}{2}-d_U}} e^{-i\pi(\frac{d+1}{2}-d_U)}  \nonumber \\
&  -\frac{g(z+p)}{(z+i\frac{\omega_n}{2}+p)^2}\frac{(z+2p)^{\frac{d+1}{2}-d_U}z^{\frac{d+1}{2}-d_U}}{(z+i\omega_n+2p)^{\frac{d+1}{2}-d_U}(z+i\omega_n)^{\frac{d+1}{2}-d_U}} e^{i\pi(\frac{d+1}{2}-d_U)} \nonumber \\
&  -\frac{g(z+p)}{(z-i\frac{\omega_n}{2}+p)^2}\frac{z^{\frac{d+1}{2}-d_U}(z+2p)^{\frac{d+1}{2}-d_U}}{(z-i\omega_n)^{\frac{d+1}{2}-d_U}(z-i\omega_n+2p)^{\frac{d+1}{2}-d_U}} e^{-i\pi(\frac{d+1}{2}-d_U)}
\Bigg). \label{eq:kd4_cut_f}
\end{align}

\subsection{$K_{para}$}
Using Eqs. \ref{eq:Kp1}, \ref{eq:Kp2}, \ref{eq:kd3_f}, \ref{eq:kd4_cut_f}, and \ref{eq:kd4_res_f}, we find
\beq
K^{ii}_{para1,n}(q) &=&  g^2\frac{\beta}{8\omega_n^2}\int \frac{d^dp}{(2\pi)^d}(2p^i+q^i)^2\csc^2\bigg(\frac{\beta}{2}(\delta+\frac{\omega_n}{2}) \bigg) \label{eq:kp1_f} \\
K^{ii}_{para2,res,n}(q) &=& \frac{g^2}{4\omega_n^2}\int \frac{d^dp}{(2\pi)^d} (2p^i+q^i)^2 \bigg( -\frac{\beta}{4}\csc^2\bigg(\frac{\beta}{2}(\delta+\frac{\omega_n}{2})\bigg) - \omega_n(\frac{d+1}{2}-d_U)\frac{\cot(\frac{\beta}{2}(\delta+\frac{\omega_n}{2}))}{p^2+(\delta+\frac{\omega_n}{2})^2}\bigg) \label{eq:kp2_res_f} 
\eeq
and
\begin{align}
K^{ii}_{para2,cut,n}(q\rightarrow0) & = g^2\frac{\sin{\pi(\frac{d+1}{2}-d_U)}}{4\pi\omega_n^2} \int \frac{d^dp}{(2\pi)^d}(2p^i+q^i)^2 \int\limits_{\varepsilon}^{\infty}dz  \nonumber \\
\Bigg(
&
\frac{g(z+p)}{(z-i\frac{\omega_n}{2}+p)^2}\frac{(z-i\omega_n+2p)^{\frac{d+1}{2}-d_U}(z-i\omega_n)^{\frac{d+1}{2}-d_U}}{(z+2p)^{\frac{d+1}{2}-d_U}z^{\frac{d+1}{2}-d_U}} e^{i\pi(\frac{d+1}{2}-d_U)} \nonumber \\
&  +\frac{g(z+p)}{(z+i\frac{\omega_n}{2}+p)^2}\frac{(z+i\omega_n)^{\frac{d+1}{2}-d_U}(z+i\omega_n+2p)^{\frac{d+1}{2}-d_U}}{z^{\frac{d+1}{2}-d_U}(z+2p)^{\frac{d+1}{2}-d_U}} e^{-i\pi(\frac{d+1}{2}-d_U)}  \nonumber \\
&  -\frac{g(z+p)}{(z+i\frac{\omega_n}{2}+p)^2}\frac{(z+2p)^{\frac{d+1}{2}-d_U}z^{\frac{d+1}{2}-d_U}}{(z+i\omega_n+2p)^{\frac{d+1}{2}-d_U}(z+i\omega_n)^{\frac{d+1}{2}-d_U}} e^{i\pi(\frac{d+1}{2}-d_U)} \nonumber \\
&  -\frac{g(z+p)}{(z-i\frac{\omega_n}{2}+p)^2}\frac{z^{\frac{d+1}{2}-d_U}(z+2p)^{\frac{d+1}{2}-d_U}}{(z-i\omega_n)^{\frac{d+1}{2}-d_U}(z-i\omega_n+2p)^{\frac{d+1}{2}-d_U}} e^{-i\pi(\frac{d+1}{2}-d_U)}
\Bigg). \label{eq:kp2_cut_f}
\end{align}
For $K_{para3}$, we rewrite Eq. \ref{eq:Kp3} in terms of $\delta$ defined in Eq. \ref{eq:delta},
\beq
K^{\mu\nu}_{para3,n}(q) &=& \frac{g^2}{4\omega_n^2}\int \frac{d^dp}{(2\pi)^d} (2p^i+q^i)^2T\sum\limits_{m}\frac{1}{( i\omega_m+i(\delta+\frac{\omega_n}{2}))^2}\Bigg(\frac{p^2+\omega_m^2}{p^2+2\omega_n\delta -(i\omega_m+i\omega_n)^2}\Bigg)^{\frac{d+1}{2}-d_U}. \nonumber
\eeq
The calculation proceeds in the same manner as in the case of $K_{dia4}$. We simply state the results here. The contribution from the residue is
\beq
K^{ii}_{para3,res,n}(q) = \frac{g^2}{4\omega_n^2}\int \frac{d^dp}{(2\pi)^d} (2p^i+q^i)^2 \bigg(-\frac{\beta}{4}\csc^2\bigg(\frac{\beta}{2}(\delta+\frac{\omega_n}{2})\bigg) + \omega_n(\frac{d+1}{2}-d_U)\frac{\cot(\frac{\beta}{2}(\delta+\frac{\omega_n}{2}))}{p^2+(\delta+\frac{\omega_n}{2})^2}\bigg) \label{eq:kp3_res_f}
\eeq
and the contribution form the branch cut (with $\frac{d-1}{2}<d_U<\frac{d+3}{2}$) is
\begin{align}
K^{ii}_{para3,cut,n}(q\rightarrow0) = & g^2\frac{\sin{\pi(\frac{d+1}{2}-d_U)}}{4\pi\omega_n^2} \int \frac{d^dp}{(2\pi)^d}(2p^i+q^i)^2 \int\limits_{\varepsilon}^{\infty}dz  \nonumber \\
\Bigg(
&
\frac{g(z+p)}{(z-i\frac{\omega_n}{2}+p)^2}\frac{(z-i\omega_n+2p)^{\frac{d+1}{2}-d_U}(z-i\omega_n)^{\frac{d+1}{2}-d_U}}{(z+2p)^{\frac{d+1}{2}-d_U}z^{\frac{d+1}{2}-d_U}} e^{i\pi(\frac{d+1}{2}-d_U)} \nonumber \\
&  +\frac{g(z+p)}{(z+i\frac{\omega_n}{2}+p)^2}\frac{(z+i\omega_n)^{\frac{d+1}{2}-d_U}(z+i\omega_n+2p)^{\frac{d+1}{2}-d_U}}{z^{\frac{d+1}{2}-d_U}(z+2p)^{\frac{d+1}{2}-d_U}} e^{-i\pi(\frac{d+1}{2}-d_U)}  \nonumber \\
&  -\frac{g(z+p)}{(z+i\frac{\omega_n}{2}+p)^2}\frac{(z+2p)^{\frac{d+1}{2}-d_U}z^{\frac{d+1}{2}-d_U}}{(z+i\omega_n+2p)^{\frac{d+1}{2}-d_U}(z+i\omega_n)^{\frac{d+1}{2}-d_U}} e^{i\pi(\frac{d+1}{2}-d_U)} \nonumber \\
&  -\frac{g(z+p)}{(z-i\frac{\omega_n}{2}+p)^2}\frac{z^{\frac{d+1}{2}-d_U}(z+2p)^{\frac{d+1}{2}-d_U}}{(z-i\omega_n)^{\frac{d+1}{2}-d_U}(z-i\omega_n+2p)^{\frac{d+1}{2}-d_U}} e^{-i\pi(\frac{d+1}{2}-d_U)}
\Bigg). \label{eq:kp3_cut_f}
\end{align}

\subsection{Total Response Function and Conductivity}
The contributions from the residues at $-i(\frac{\omega_n}{2}+\delta)$ vanish for both diamagnetic and paramagnetic terms, separately. To check this, we add the contributions from the residue at $-i(\frac{\omega_n}{2}+\delta)$ from $K_{dia2}$, $K_{dia3}$, and $K_{dia4,res}$,
\begin{align}
\int \frac{d^dp}{(2\pi)^d}(2p^i+q^i)^2  &\bigg(-\frac{1}{2\omega_n}(\frac{d+1}{2}-d_U)\frac{\cot(\frac{\beta}{2}(\delta+\frac{\omega}{2}))}{p^2+(\delta+\frac{\omega_n}{2})^2} -\frac{\beta}{8\omega^2_n}\csc^2\bigg(\frac{\beta}{2}(\delta+\frac{\omega_n}{2})\bigg) \nonumber \\
&+\frac{\beta}{8\omega^2_n}\csc^2\bigg(\frac{\beta}{2}(\delta+\frac{\omega_n}{2})\bigg) + \frac{1}{2\omega_n}(\frac{d+1}{2}-d_U)\frac{\cot(\frac{\beta}{2}(\delta+\frac{\omega_n}{2}))}{p^2+(\delta+\frac{\omega_n}{2})^2}
\bigg) \nonumber \\
= & 0, \nonumber
\end{align}
and add the contrubutions from the residue at $-i(\frac{\omega_n}{2}+\delta)$ from $K_{para1}$, $K_{para2}$, and $K_{para3}$,
\begin{align}
\int \frac{d^dp}{(2\pi)^d}(2p^i+q^i)^2 &
\bigg(\frac{\beta}{8\omega^2_n}\csc^2\bigg(\frac{\beta}{2}(\delta+\frac{\omega_n}{2})\bigg) \nonumber \\
&-\frac{\beta}{16\omega^2_n}\csc^2\bigg(\frac{\beta}{2}(\delta+\frac{\omega_n}{2})\bigg) - \frac{1}{4\omega_n}(\frac{d+1}{2}-d_U)\frac{\cot(\frac{\beta}{2}(\delta+\frac{\omega_n}{2}))}{p^2+(\delta+\frac{\omega_n}{2})^2} \nonumber \\
&-\frac{\beta}{16\omega^2_n}\csc^2\bigg(\frac{\beta}{2}(\delta+\frac{\omega_n}{2})\bigg) + \frac{1}{4\omega_n}(\frac{d+1}{2}-d_U)\frac{\cot(\frac{\beta}{2}(\delta+\frac{\omega_n}{2}))}{p^2+(\delta+\frac{\omega_n}{2})^2} \nonumber \\
= & 0. \nonumber
\end{align}
We can safely take the limit $q\rightarrow0$ and $\delta\rightarrow0$. Futhermore, the contributions from the branch cuts, when adding both paramagnetic ($K_{para2}$ and  $K_{para3}$) and diamagnetic terms ($K_{dia4}$),  vanish, i.e. Eq. \ref{eq:kd4_cut_f} + Eq. \ref{eq:kp2_cut_f} + Eq. \ref{eq:kp3_cut_f} is equal to zero. The total optical conductivity in Matsubara space is
\beq \label{eq:final_conductivity}
\sigma^{ii}(i\omega_n) &=& 2g^2(\frac{d+1}{2}-d_U)\int \frac{d^dp}{(2\pi)^d} p^2_i \Biggl\{  \frac{\beta}{\omega_n p^2}N(p)(N(p)+1) + \frac{ig(p)}{2p^3}\bigg(\frac{1}{2p+i\omega_n}-\frac{1}{2p-i\omega_n} \bigg) \Biggl\} \nonumber \\
&=& (\frac{d+1}{2}-d_U)\sigma^{ii}_0(i\omega_n)  
\eeq
where $\sigma^{ii}_0$ is optical conductivity of a free scalar field (Eq. \ref{eq:massive_scalar_conductivity} with m = 0).

\bibliography{mybib}
\bibliographystyle{apsrev4-1}
\end{document}